\documentclass[reprint,twocolumn,superscriptaddress,numerical,pra,floats]{revtex4-2}

\usepackage[T1]{fontenc}
\usepackage[utf8]{inputenc}
\usepackage{bbold}
\usepackage{amsmath}
\usepackage{bm}
\usepackage{physics}
\usepackage{calrsfs}
\usepackage{graphicx}
\usepackage{bbm}
\usepackage{xcolor}
\usepackage{xr-hyper}
\usepackage{hyperref}
\DeclareMathAlphabet{\pazocal}{OMS}{zplm}{m}{n}

\newcommand{\defeq}{\mathrel{\llap{$:$}}=}
\usepackage[protrusion=true, expansion=true,
tracking=true, kerning=true, spacing=true,
final]{microtype}
\usepackage{hyphenat}
\usepackage{enumitem}
\usepackage{booktabs}
\usepackage{tikz}
\usetikzlibrary{calc,arrows.meta}
\usepackage{pgfplots}
\pgfplotsset{compat=1.17}
\usetikzlibrary{decorations.markings}
\tikzset{
    set arrow inside/.code={\pgfqkeys{/tikz/arrow inside}{#1}},
    set arrow inside={end/.initial=stealth, opt/.initial=},
    /pgf/decoration/Mark/.style={
        mark/.expanded=at position #1 with
        {
            \noexpand\arrow[\pgfkeysvalueof{/tikz/arrow inside/opt}]{\pgfkeysvalueof{/tikz/arrow inside/end}}
        }
    },
    arrow inside/.style 2 args={
        set arrow inside={#1},
        postaction={
            decorate,decoration={
                markings,Mark/.list={#2}
            }
        }
    },
}

\makeatletter
\newcommand{\supplementarytoc}{%
    \begingroup
        \@ifundefined{toc@pre@auto}{}{%
            \let\toc@pre\toc@pre@auto
            \let\toc@post\toc@post@auto
        }%
        \let\l@title\@gobbletwo
        \let\l@abstract\@gobbletwo
        \let\l@section\@gobbletwo
        \@starttoc{toc}%
    \endgroup
}
\newcommand{\numbersupplementary}{%
    \setcounter{secnumdepth}{3}%
    \setcounter{subsection}{0}%
    \def\thesubsection{S\arabic{subsection}}%
    \def\thesubsubsection{\thesubsection.\arabic{subsubsection}}%
    \def\p@subsection{}%
    \def\p@subsubsection{}%
}
\makeatother

\begin{document}
\setcounter{topnumber}{1}

\title{Scalable simulation of strongly interacting trapped ultracold bosons with an exact pair-correlated ansatz}

\author{Dawid A. Hryniuk}
\email{d.hryniuk@ucl.ac.uk}
\affiliation{Department of Physics and Astronomy, University College London,
Gower Street, London, WC1E 6BT, United Kingdom}
\affiliation{Center for Theoretical Physics, Polish Academy of Sciences, Al. Lotników 32/46, 02-668 Warsaw, Poland}
\author{Mateusz Pańtak}
\affiliation{Center for Theoretical Physics, Polish Academy of Sciences, Al. Lotników 32/46, 02-668 Warsaw, Poland}
\author{Kazimierz Rzążewski}
\affiliation{Center for Theoretical Physics, Polish Academy of Sciences, Al. Lotników 32/46, 02-668 Warsaw, Poland}

\date{\today}

\begin{abstract}

    Cold-atom experiments routinely realize strongly correlated phases with tunable interactions and trapping potentials, demanding scalable simulation methods which stay accurate in these regimes.
    We introduce a variational Monte Carlo method with a pair-correlated ansatz built from the exact harmonically trapped two-body solution, which we comprehensively benchmark against exact diagonalization, analytical, and mean-field results.
    Reaching up to $10^5$ atoms, it accurately captures the ground-state properties of the gas across a broad range of interaction strengths and confinements.
    With only two independent variational parameters, it quantifies the condensate depletion and entanglement entropies throughout the full crossover from an ideal condensate to a fermionized Tonks--Girardeau gas of $10^3$ atoms in 1D.
    Our method opens new routes towards scalable simulation of ultracold atomic and molecular gases and mixtures.
\end{abstract}

\maketitle

\section{Introduction}

Over the past three decades, ultracold atomic gases have matured into a versatile platform for the study of quantum many-body physics, with applications ranging from the simulation of condensed-matter~\cite{Bloch2008} and nuclear systems~\cite{Hammer2013} to cold chemistry~\cite{Zhang2023,Bohn2017} and quantum metrology~\cite{Huang2024,Pezze2018}.
This has been enabled by experimental advances providing exceptional control over atom numbers~\cite{Kaufman2021,Serwane2011}, trapping geometry~\cite{Navon2021}, and, via magnetic and optical Feshbach resonances~\cite{RevModPhys.82.1225}, interatomic interactions.
This control has in turn revealed strongly correlated phases with no mean-field description, including confinement-induced resonances~\cite{Haller2010,Olshanii1998}, quantum droplets~\cite{Bottcher2021,Semeghini2018,doi:10.1126/science.aao5686,Petrov2015}, bound trimers~\cite{Chuang2025,Xie2020,Greene2017,Kraemer2006,Efimov1970}, Tonks--Girardeau~\cite{Malvania2021,Kao2021,Wilson2020,Kinoshita2004,Paredes2004} and unitary Bose gases~\cite{Wang2025,Eigen2018,Fletcher2017,Makotyn2014}.
Quantitative description of these regimes rests on numerical methods capable of accurately capturing the many-body correlations measurable in modern experiments~\cite{Yao2025,Xiang2025,Wang2026}.

At low temperatures and weak inter-atomic repulsion, trapped Bose gases are well described by the Gross--Pitaevskii equation~\cite{Dalfovo1999,Gross1961,Pitaevskii1961}, together with Bogoliubov~\cite{Bogoliubov1947} and Lee--Huang--Yang~\cite{Skov2021,PhysRev.106.1135} corrections; in 1D, they are also exactly solvable at infinite repulsion via Bose--Fermi mapping~\cite{Girardeau2001,Girardeau1960}.
At large atom numbers and intermediate coupling, where correlations become important and most tunable-interaction experiments operate, quantum Monte Carlo is a standard non-perturbative approach~\cite{Ceperley1995}.
Path-integral Monte Carlo typically handles thousands of atoms~\cite{Boninsegni2006,Krauth1996}, but reaches the ground state only by extrapolating to low temperature, requiring many time slices, with the computational cost further compounded at strong coupling~\cite{Sakkos2009}.
Diffusion Monte Carlo works directly at zero temperature and is in principle exact for nodeless bosonic ground states~\cite{Giorgini1999,Boronat1994,PhysRevA.63.063601}.
In practice, convergence requires control of finite-time-step and walker-population errors.
Moreover, its standard formulation produces mixed estimates for many observables, including the one-body density matrix~\cite{DuBois2003}.

Variational Monte Carlo (VMC) provides a complementary approach based on an explicit parameterized many-body wavefunction.
Pair-correlated VMC states have been applied across broad interaction regimes~\cite{Astrakharchik2002,PhysRevA.63.023602}, primarily in 3D traps, while their 1D counterparts have largely been restricted to smaller systems~\cite{PhysRevLett.108.045301,Brouzos2013} or used as guiding functions for projector methods~\cite{Astrakharchik2005,Astrakharchik2003}.
More recent neural-network~\cite{PhysRevC.109.034004,Pescia2022,Pescia2024} and tensor-network~\cite{PhysRevA.75.021601,PhysRevLett.128.020501,Dutta2022,Verstraete2010} representations promise greater expressivity, though often at the cost of many thousands of parameters, demanding optimization, and limited system sizes.

To overcome these limitations, we introduce a scalable VMC method for characterizing the strongly correlated ground state properties of harmonically trapped Bose gases with tunable interactions and confinement geometries, based on the exact solution to the trapped two-body problem.
Requiring as few as two independent variational parameters, its low optimization cost lets us target experimentally realistic system sizes of up to $N=10^5$ atoms.
By benchmarking against analytical, exact-diagonalization, exact Tonks--Girardeau, and mean-field and local-density results, we show that a two-body-correlated trial state captures the ground-state energy to within $0.15\%$ for few atoms and $1.8\%$ for $10^3$ atoms across the full crossover from an ideal to a fermionized Tonks--Girardeau gas in 1D.

\begin{figure*}
    \includegraphics[width=\textwidth]{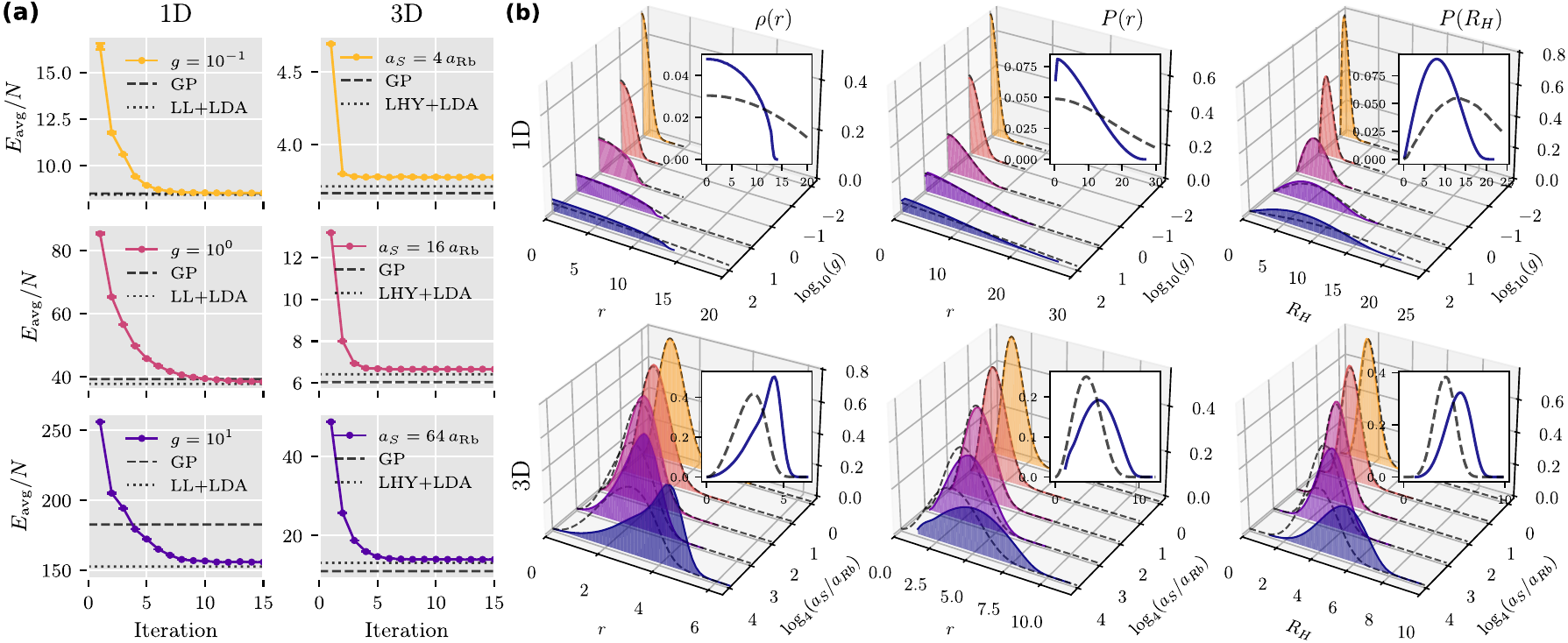}
    \caption{(a) VMC energy per atom over stochastic-reconfiguration iterations for $N=1000$ atoms in 1D (left) and 3D (right) isotropic traps for several $g$ and $a_s$, compared with the Gross-Pitaevskii (GP) energy and with GP including Lieb-Liniger (LL, 1D) or Lee-Huang-Yang (LHY, 3D) corrections in a local density approximation (LDA).
    (b) Distributions of atom--trap-centre distances $\rho(r)$, pair distances $P(r)$, and triplet hyperradii $P(R_H)$ for $N=100$ in 1D (top) and 3D (bottom), compared with GP (grey dashed lines).
    Insets: strongest interactions ($g=100$, $a_s = 256a_{\text{Rb}}$).}
    \label{fig: histograms}
\end{figure*}

\section{Hamiltonian and variational ansatz}

We simulate the ground state of $N$ identical bosons of mass $m$ with two-body interactions in 1D and 3D harmonic traps, with Hamiltonian
\begin{equation}
    H = \sum_{i=1}^{N} \left( -\frac{\hbar^2}{2 m} \nabla_{\vb{r}_i}^2 + \frac{m}{2} \vb{r}_i^\top \Omega^2 \vb{r}_i \right) + \sum_{i<j} V(\vb{r}_i - \vb{r}_j),
    \label{eq: Hamiltonian}
\end{equation}
where $\vb{r}_i = \text{r}_{i}^{n} \in \mathbb{R}^d$, $n=\{x,y,z\}$, denotes the position of the $i^{\text{th}}$ atom relative to the centre of the trap, and $d$ is the number of spatial dimensions.
In the most general case, we will consider a 3D anisotropic harmonic trapping potential with $\Omega = \text{diag}(\omega_x, \omega_y, \omega_z)$.
In 1D, we will assume the atoms to be point-like with contact interaction of the form $V(r) = g\delta(r)$, where $g$ is the interaction strength and $\delta(r)$ denotes the Dirac delta function, while in 3D we will model the system as an ensemble of hard spheres with $s$-wave scattering length $a_s$, i.e., $V(\bm{r})=\infty$ for $|\bm{r}|\leq a_s$ and $V(\bm{r})=0$ otherwise.
While a 3D gas is often also modelled as point particles interacting via a Fermi pseudopotential, the resultant Hamiltonian is not self-adjoint for more than two particles and is thus not amenable to the variational treatment discussed in this work \cite{Huang_1987,Braaten2006}.

The Hamiltonian \eqref{eq: Hamiltonian} is separable into centre-of-mass and relative parts when transforming to the respective coordinates (see Supplementary Sec.~\ref{supp: separation}),
\begin{align}
    H_{\text{com}} &= \frac{1}{2Nm}\vb{P}^2 + \frac{Nm}{2} \vb{X}^\top \Omega^2 \vb{X}, \label{eq: H_com}\\
    \begin{split}
        H_{\text{rel}} &= -\frac{\hbar^2}{2m} \sum_{i<j, k<l} (\delta_{ik}-\delta_{il}-\delta_{jk}+\delta_{jl}) \nabla_{\bm{r}_{ij}} \!\cdot\! \nabla_{\bm{r}_{kl}} \\ &\qquad + \frac{m}{2N} \sum_{i<j}\bm{r}^\top_{ij} \Omega^2 \bm{r}_{ij} + \sum_{i<j} V(\bm{r}_{ij}),
    \end{split} \label{eq: H_rel}
\end{align}
where $\vb{X}=\frac{1}{N}\sum_{i}\vb{r}_i$ is the centre-of-mass coordinate, $\vb{P} = -i\hbar\sum_i \nabla_{\vb{r}_i}$ is the total momentum, $\bm{r}_{ij}=\vb{r}_i-\vb{r}_j$, and $\delta_{ij}$ denotes the Kronecker delta.
For $N=2$ in isotropic traps of arbitrary dimension, the exact ground state of Eq.~\eqref{eq: Hamiltonian} is the product of the ground states of $H_{\text{com}}$ and $H_{\text{rel}}$~\cite{Busch1998}.
Since the contact interactions considered here are pairwise, this exact solution motivates a variational ansatz for the (unnormalized) $N$-body ground-state wavefunction:
\begin{equation}
    \Psi(\vb{R}; \bm{\theta}) = \Phi(\vb{X}) \prod_{i<j} f(\bm{r}_{ij}; \bm{\theta}),
    \label{eq: trial state}
\end{equation}
where $\vb{R}=(\vb{r}_1,\ldots,\vb{r}_N)$ and the product runs over all distinct pairs of atoms.
Here, $\Phi(\vb{X})$ is the centre-of-mass ground state, and the relative part $\Psi_{\mathrm{rel}}(\vb{R};\bm{\theta})\equiv\prod_{i<j}f(\bm{r}_{ij};\bm{\theta})$ is a product of variationally parameterized pair factors.
For an isotropic trap in $d$ spatial dimensions, the pair factors take the two-parameter form
\begin{equation}
    f(\bm{r}; \alpha, \beta) = U\left(-\nu_{\alpha}, \frac{d}{2}, \alpha^2 \tilde{r}^2\right) \exp\left(-\frac{\beta}{2} \tilde{r}^2\right),
    \label{eq: pair product}
\end{equation}
where $U$ denotes Tricomi's confluent hypergeometric function, $\tilde{r}^2=r^2/a_0^2$, $r=\lVert\bm{r}\rVert$, and $a_0=\sqrt{\hbar/(\mu\omega)}$ is the relative-coordinate oscillator length, with reduced mass $\mu=m/2$, so that the single-particle oscillator length is $a_{\mathrm{ho}}=a_0/\sqrt{2}$.
For hard spheres of diameter $a_s$, the pair factor is additionally made to vanish for $r<a_s$.
In comparison to the more commonly encountered Jastrow--Bijl form, the ansatz~\eqref{eq: trial state} makes use of a single non-parameterized centre-of-mass factor $\Phi(\vb{X})$, which, for harmonic confinement, can equivalently be factorized into one- and two-body Gaussian factors.

For $\alpha=\beta=1$, Eq.~\eqref{eq: trial state} reduces to the exact $N=2$ solution for point particles in a $d$-dimensional isotropic trap \cite{Busch1998} and for hard spheres in 3D; for $N>2$, $\alpha$ and $\beta$ intuitively account for effective changes in the atoms' spatial distribution.
The number $\nu_{\alpha}$ is a root of a transcendental equation which must be solved self-consistently for each value of $\alpha$; in 1D, the equation in question arises from imposing the Lieb--Liniger cusp condition ${f'(r=0^+)} = \frac{\mu g}{\hbar^2} f(r=0)$ \cite{PhysRev.130.1605} on the pair factors and takes the form
\begin{equation}
    \alpha \Gamma(-\nu_\alpha+\tfrac{1}{2}) + \frac{g}{2\hbar\omega a_0}\Gamma(-\nu_\alpha) = 0,
\end{equation}
where $\Gamma$ denotes the gamma function.
We take the root on the branch continuously connected to the non-interacting solution $\nu_\alpha=0$ at $g=0$, i.e., $\nu_\alpha\in(0,\tfrac12)$ for $g>0$ and $\nu_\alpha<0$ for $g<0$.
For hard spheres, one instead enforces the boundary condition $f(|\bm{r}| = a_s)=0$, leading to
\begin{equation}
    U\left(-\nu_{\alpha}, \frac{d}{2}, \frac{\alpha^2 a_s^2}{a_0^2}\right) = 0,
    \label{eq: hard sphere condition}
\end{equation}
where $\nu_\alpha$ is the smallest positive root, for which $f$ has no node for $r>a_s$.

In 1D, the ansatz is exact in both limits of the coupling $g$.
At $g=0$, $\nu_\alpha=0$ and $U=1$, so Eq.~\eqref{eq: trial state} with $\beta=2/N$ reproduces the noninteracting product state.
As $g\to+\infty$, $\nu_\alpha\to\tfrac{1}{2}$ and $U(-\tfrac{1}{2},\tfrac{1}{2},z)=\sqrt{z}$, so the same expression reduces to $\prod_{i<j}|x_i-x_j|e^{-\sum_i x_i^2/2}$, i.e. the Tonks--Girardeau (TG) ground state \cite{Girardeau2001}.

The above construction generalizes straightforwardly to $d$-dimensional \textit{anisotropic} harmonic traps, with independent Gaussian parameters $\beta_n$ along each trap axis, where we write
\begin{equation}
    f(\bm{r};\alpha,\bm{\beta})
    =
    U\!\left(
    -\nu_\alpha,\frac{d}{2},
    \alpha^2\tilde{r}^2
    \right)
    \exp\!\left(
    -\frac{1}{2}\sum_{n=1}^{d}
    \beta_n\tilde{r}_n^2
    \right),
    \label{eq: pair product anisotropic}
\end{equation}
containing $d+1$ variational parameters, $\bm{\theta}=(\alpha,\beta_1,\ldots,\beta_d)$, where $\tilde r_n\equiv r_n/a_n$, $a_n^2=\hbar/(\mu\omega_n)$, and $a_0^2=d^{-1}\sum_{n=1}^{d}a_n^2$.
The single parameter $\alpha$ controls the radial Tricomi factor, while the parameters $\beta_n$ allow the Gaussian envelope to respond independently to the anisotropic confinement along each spatial direction.
Since the Gaussian factors are nonzero at finite separation, the spherical hard-sphere boundary is determined solely by the direction-independent radial Tricomi factor, and the value of $\nu$ is again imposed by Eq.~\eqref{eq: hard sphere condition}.

The 1D version of Eq.~\eqref{eq: pair product} involving parabolic cylinder functions was explored in earlier works as an analytical ansatz for the 1D Bose gas in single harmonic and periodic potentials \cite{PhysRevLett.108.045301, Brouzos2013}.
Related variational wavefunctions for few atoms in 1D interpolate between the non-interacting and infinitely repulsive limits~\cite{Andersen2016} or employ optimized pair-correlation functions in arbitrary traps~\cite{Koscik2018,Koscik2021}.
For 3D Bose gases in isotropic and anisotropic traps, earlier works \cite{PhysRevA.63.023602,DuBois2003,Astrakharchik2002,Astrakharchik2004Q1D} used hard spheres with Jastrow trial states with parameterized one- and parameter-free two-body factors, the latter of the form $f(r) = 1-\frac{a_s}{r}$ for $r>a_s$, i.e. the zero-energy $s$-wave scattering solution outside the hard sphere in free space; this expression does not, in general, match the corresponding exact solution for harmonically trapped atoms in Eq.~\eqref{eq: pair product} (for $\alpha=\beta=1$).
In this work, we formulate a new efficient variational Monte Carlo method based on the trial state \eqref{eq: trial state} with the parameterized pair-product factors \eqref{eq: pair product} and \eqref{eq: pair product anisotropic} and applicable to the quantum many-body problem in 1D and 3D isotropic and anisotropic harmonic traps, targeting large systems in strongly correlated regimes.

\section{Variational Monte Carlo}

To find the optimal ground state parameter values $\bm{\theta}$, we resort to the usual time-independent variational principle, minimizing the energy functional
\begin{equation}
    E_{\bm{\theta}}[\Psi] = \frac{\int \dd{\vb{R}}\,\Psi^*(\vb{R}; \bm{\theta})\,H\,\Psi(\vb{R}; \bm{\theta})}{\int \dd{\vb{R}}\,|\Psi(\vb{R}; \bm{\theta})|^2}.
\end{equation}
Its exact evaluation is not possible in practice for the larger systems; instead, it can be efficiently estimated via Monte Carlo sampling.
Specifically, we recast the expression for the energy as an ensemble average over many-body configurations $\{\vb{R}_k\}_{k=1}^{N_{\text{MC}}}$ drawn from the Born distribution $\rho(\vb{R}; \bm{\theta}) = |\Psi(\vb{R}; \bm{\theta})|^2/\int \dd{\vb{R}}\,|\Psi(\vb{R}; \bm{\theta})|^2$,
\begin{subequations}\label{eq: local energy 1}
    \renewcommand{\theequation}{\theparentequation.\arabic{equation}}
    \begin{align}
        E_{\bm{\theta}}[\Psi] &= \mathbbm{E}_{\vb{R}\sim \rho(\vb{R}; \bm{\theta})}[E_\text{loc}(\vb{R}; \bm{\theta})] \\
        &\defeq \langle E_\text{loc}(\vb{R}; \bm{\theta})\rangle \\
        &\approx \frac{1}{N_\text{MC}} \sum_{k=1}^{N_\text{MC}}  \frac{H\Psi(\vb{R}_k; \bm{\theta})}{\Psi(\vb{R}_k; \bm{\theta})}
    \end{align}
\end{subequations}
where $E_\text{loc}(\vb{R};\bm{\theta}) = H\Psi(\vb{R}; \bm{\theta})/\Psi(\vb{R}; \bm{\theta})$ is usually referred to as the local energy.
Noting the earlier discussed decomposition and suppressing the common arguments of $\Psi_{\mathrm{rel}}$, it simplifies into the sum
\begin{equation}
    E_\text{loc}(\vb{R};\bm{\theta}) = \frac{T\Psi_{\mathrm{rel}}}{\Psi_{\mathrm{rel}}} + \frac{V_{\mathrm{trap}}\Psi_{\mathrm{rel}}}{\Psi_{\mathrm{rel}}} + E_{\text{com}},
    \label{eq: local energy 2}
\end{equation}
where $T$ and $V_{\mathrm{trap}}$ denote the kinetic and trapping-potential energy operators of the relative Hamiltonian, respectively, while $E_{\text{com}}$ is the ground-state energy of the centre-of-mass part.
No interaction-energy term enters the local energy by virtue of the contact conditions imposed on $f$.
Upon changing to relative coordinates, the potential-energy term becomes
\begin{equation}
    \frac{V_{\mathrm{trap}}\Psi_{\mathrm{rel}}}{\Psi_{\mathrm{rel}}} = \frac{\mu}{N} \sum_{i<j} \bm{r}_{ij}^\top \Omega^2 \bm{r}_{ij},
\end{equation}
while the radial Laplacian takes the general form (see Supplementary Sec. \ref{supp: kinetic} for derivation):
\begin{equation}
    \begin{aligned}
        \frac{T\Psi_{\mathrm{rel}}}{\Psi_{\mathrm{rel}}} &= -\frac{\hbar^2}{2\mu} \sum_{i<j} \Bigg[ \frac{\nabla^2_{\bm{r}_{ij}} f(\bm{r}_{ij}; \bm{\theta})}{f(\bm{r}_{ij}; \bm{\theta})}
        - \left\|\nabla_{\bm{r}_{ij}}\ln f(\bm{r}_{ij}; \bm{\theta}) \right\|^2  \Bigg]
        \\ &\qquad - \frac{\hbar^2}{4\mu}  \|\bm{v}(\vb{R}; \bm{\theta})\|^2, %
    \end{aligned}
    \label{eq: kinetic energy}
\end{equation}
where we defined the vector $ \bm{v}_k(\vb{R};\bm{\theta}) = \sum_{i<j}(\delta_{ik}-\delta_{jk})\nabla_{\bm{r}_{ij}}\ln f(\bm{r}_{ij};\bm{\theta}). $

To find the optimal parameter values minimizing the energy efficiently, we make use of the stochastic reconfiguration method, which amounts to repeatedly solving the set of linear equations
\begin{equation}
    \sum_{j} S_{ij}\delta \theta_j = F_i
\end{equation}
where, suppressing the arguments $(\vb{R};\bm{\theta})$, we defined
\begin{align}
    F_i &= \tfrac12 \partial_{\theta_i} E_{\bm{\theta}}[\Psi] = \langle E_\text{loc} \Delta_i \rangle - \langle E_\text{loc} \rangle \langle \Delta_i \rangle, \\
    S_{ij} &= \langle \Delta^\dagger_i \Delta_j \rangle - \langle \Delta^\dagger_i \rangle \langle \Delta_j \rangle,
\end{align}
and the vector of log-derivatives $\Delta_i(\vb{R}; \bm{\theta}) = \partial_{\theta_i} \ln \Psi(\vb{R}; \bm{\theta})$.
In practice, the metric tensor is ensured to be invertible by means of Tikhonov regularization $S_{ij} \rightarrow S_{ij} + c\delta_{ij}$, where a constant $c \simeq 10^{-3}$ was found to be optimal for rapid and robust convergence.
In contrast to simple stochastic gradient descent (corresponding to $S_{ij} = \delta_{ij}$), the stochastic reconfiguration method takes into account the geometry of the variational manifold encoded by the metric tensor $S_{ij}$, leading to more efficient optimization and faster convergence to the optimal parameter values \cite{Sorella2007}.
The variational parameters are then updated as $\bm{\theta} \leftarrow \bm{\theta} - \eta\,\delta\bm{\theta}$, where $\eta$ is a small step size controlling the rate of convergence.

To generate the set of Monte Carlo samples $\{\vb{R}_k\}_{k=1}^{N_{\text{MC}}}$ required to evaluate ensemble averages, we make use of simple local Metropolis updates, shifting a randomly selected atom $\vb{r}_k \in \vb{R}$ to a neighbouring position $\vb{r}_k' \in \vb{R}'$ drawn from a uniform proposal density centred on $\vb{r}_k$, with acceptance probability $\min\{1,|\Psi(\vb{R}'; \bm{\theta})|^2/|\Psi(\vb{R}; \bm{\theta})|^2\}$.
Since only the pair factors involving the displaced atom change, all remaining ratios cancel and the acceptance probability is evaluated efficiently in $O(N)$ time.
The width of the proposal density is adapted on the fly to maintain a mean acceptance rate of around $30\%$ for optimal sample decorrelation.
Sample autocorrelation is monitored throughout the optimization via the integrated autocorrelation time $\tau$ of the local energy; the gradient contributions are accumulated only every $\kappa^{\text{th}}$ configuration, with the thinning interval $\kappa$ adjusted dynamically so that $\tau$ remains close to unity, and any residual correlations are accounted for in the reported Monte Carlo standard errors.
Finally, we make use of MPI to efficiently parallelize the sampling of new particle configurations, with the final configurations of each optimization step reused, after a short burn-in, as the starting points of the next.
See Supplementary Sec. \ref{supp: metropolis} for a more comprehensive discussion of sampling.

The dominant computational cost of the algorithm lies in the repeated evaluation of the confluent hypergeometric functions during sampling and optimization, which becomes prohibitive as the number of atoms and Monte Carlo samples grows.
To address this, we replace the functions by interpolants constructed on a grid linear in $\sqrt{z}$ at fixed grid-spacing density, rebuilt at the start of each optimization step over a range inferred from the spatial extent of the preceding batch of samples and enlarged by a safety factor.
This reduces the cost of evaluation to an $O(1)$ lookup, at a discretization error which we verify to lie below the Monte Carlo statistical error.
For a comprehensive exposition of the method and convergence checks with respect to the interpolant discretization, see Supplementary Sec. \ref{supp: interpolants}.

\begin{figure}[t!]
    \includegraphics[width=\columnwidth]{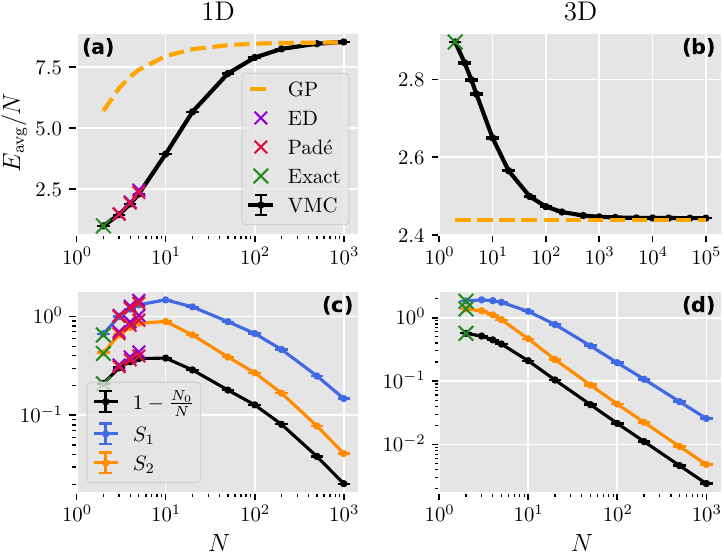}
    \caption{ Ground-state properties of the repulsive trapped Bose gas versus the number of atoms $N$.
    (a) Energy per atom in 1D with $g=100/N$, compared with GP, exact diagonalization (ED), Padé-extrapolated ED, and the exact result for $N=2$.
    (b) As in (a) for 3D hard spheres with $a_s = 1024 a_{\text{Rb}}/N$.
    (c) Depletion and von Neumann and R\'enyi-2 entanglement entropies for the gas in (a); (d) the same for (b).
    Points in (c-d) are means over 5 independent sampling runs at the same optimized parameters; error bars show the standard deviation across these runs and lie within the markers.
    ED details: Supplementary Sec. \ref{supp: ED}.}
    \label{fig: large N limit}
\end{figure}

\section{Numerical results}

From now on, we will work in units of $\hbar = m = \omega = 1$.
We first test our method in 1D and 3D isotropic traps (Fig.~\ref{fig: histograms}), comparing ground-state energies and selected distributions with mean-field Gross--Pitaevskii (GP) predictions across a wide range of interaction strengths.
The GP ground states are found by numerically integrating in imaginary time the time-dependent GP equation
\begin{equation}
    i\hbar \partial_t \phi(\vb{r}, t) = \left( -\frac{\hbar^2}{2m} \nabla^2 + \frac{m}{2} \vb{r}^\top \Omega^2 \vb{r} + g|\phi(\vb{r}, t)|^2 \right) \phi(\vb{r}, t),
    \label{eq: GP equation}
\end{equation}
with the condensate wavefunction normalized as $\int \dd{\vb{r}}\,|\phi(\vb{r},t)|^2=N$.
Throughout, $g:=\frac{4\pi \hbar^2 a_s}{m}$ in 3D, with $a_s$ expressed in multiples of $a_{\text{Rb}}=0.00433\,a_{\text{ho}}$, a commonly used $s$-wave scattering length for ultracold $^{87}$Rb atoms \cite{PhysRevA.63.023602}.
Fig.~\ref{fig: histograms} (a) shows example stochastic-reconfiguration energy minimizations for $N=1000$ atoms, compared with the converged GP ground-state energy and with GP corrected for quantum fluctuations via the Lieb--Liniger (LL, 1D) \cite{PhysRev.130.1605} or Lee--Huang--Yang (LHY, 3D) \cite{PhysRev.106.1135} corrections to the energy density in Eq. \eqref{eq: GP equation} in a local density approximation (LDA)~\cite{Dunjko2001}.
In 1D, the converged VMC ground state energies agree closely with GP at weak interactions and fall below it at stronger interactions, approaching the LL+LDA predictions.
In 3D, the VMC energies exceed GP, in agreement with earlier observations \cite{PhysRevA.63.023602}, where this was attributed to higher-energy non-condensed atoms in the hard-sphere model.

In Fig. \ref{fig: histograms} (b), we study the normalized distributions $\rho(r)$, $P(r)$, and $P(R_H)$ of the atom--trap-centre distances $|\vb{r}_i|$, pair distances $|\bm{r}_{ij}|$, and triplet hyperradii $R_{H,ijk}$, respectively, where $R_{H,ijk}^2=\frac{1}{3}\left(|\bm{r}_{ij}|^2+|\bm{r}_{ik}|^2+|\bm{r}_{jk}|^2\right)$.
These were sampled from the $N=100$ ground state and compared with GP.
These distributions are commonly measured in experiment \cite{Kinoshita2004,Yao2025,Wang2026}, the hyperradii being relevant to Efimov trimers \cite{doi:10.1126/science.aaa5601} and three-body losses \cite{LaburtheTolra2004}.
All are in excellent agreement with GP at weak interactions but deviate significantly at stronger ones, owing to correlations captured by our variational wavefunction.
Mirroring the energy, all three distributions shift towards larger distances in 3D, where the VMC energies exceed GP, and towards smaller distances in 1D, where they fall below it.
In 1D, $P(r)$ also shows an antibunching hole at small $r$, characteristic of the fermionic nature of strongly repulsive bosons.

\begin{figure}[t!]
    \includegraphics[width=\columnwidth]{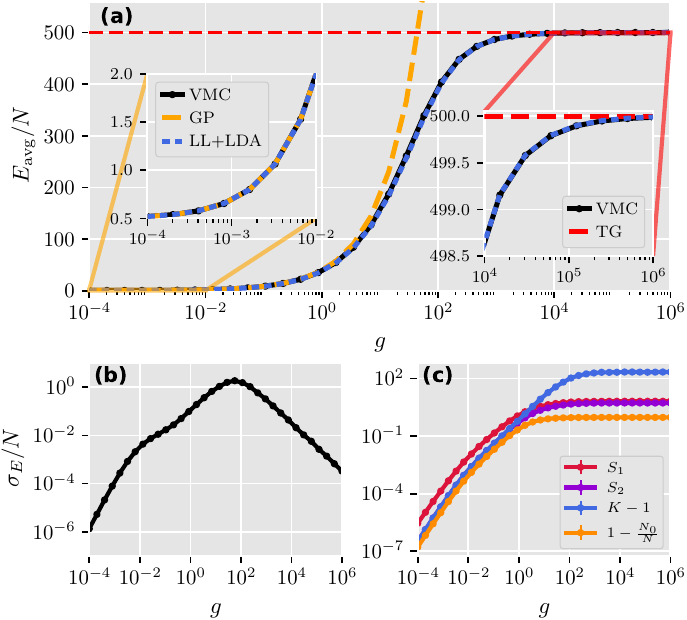}
    \caption{Repulsive 1D trapped gas of $N=1000$ atoms versus the interaction strength $g$.
    (a) Energy per atom compared with GP, LL+LDA, and the TG energy, $E_{\text{TG}} = N^2/2$; insets: very weak and very strong interactions.
    (b) Standard deviation $\sigma_E$ of the sampled local energies, per atom.
    (c) OBDM correlation measures: entropies $S_1$ and $S_2$, effective Schmidt rank minus one $K-1$, and depletion $1-N_0/N$.
    Error bars in (c), from jackknife resampling over 8 Markov chains, lie within the markers.}
    \label{fig: large g limit}
\end{figure}

It was formally shown that, in the limit $N\rightarrow \infty$ at fixed $N a_s$, the 3D trapped Bose gas becomes a pure condensate exactly described by mean-field GP theory \cite{PhysRevLett.88.170409, PhysRevA.61.043602}.
In Fig.~\ref{fig: large N limit} (a-b), we test this in 3D, and analogously in 1D at fixed $Ng$~\cite{LiebSeiringerYngvason2003,LiebSeiringerYngvason2004}, by comparing the VMC energy per atom with GP for up to $N=10^5$ (3D) and $N=10^3$ (1D) atoms.
At the smallest atom numbers, we further benchmark against exact diagonalization (ED) for up to $N=5$ in 1D and, at $N=2$, against the analytic two-body solutions \cite{Busch1998} in 1D and 3D, which our ansatz reproduces exactly.
For $3\le N\le5$, the VMC energies lie $1.8$--$6.5\%$ below the ED energies at every accessible basis cutoff, and agree with their extrapolations to the infinite-basis limit to within $0.15\%$ (Supplementary Sec.~\ref{supp: few-body}).
As VMC and finite-basis ED energies are both variational upper bounds, our two-parameter trial state thus bounds the ground-state energy more tightly than ED in bases of up to $1.2\times10^6$ states.
This is because the pair factor satisfies the contact condition exactly, whereas the contact cusp limits the convergence of ED to $n_{\max}^{-1/2}$ in the basis cutoff~\cite{Ernst2011,Jeszenszki2018}; moreover, the cost of VMC grows only as $O(N^2)$ per sample, compared with the combinatorial growth of the ED basis.
As expected, the VMC energies converge to GP with increasing $N$ in both geometries, from below in 1D and from above in 3D, consistent with the negative LL and positive LHY corrections discussed above.

To quantify the quantum correlations in the gas, we compute and diagonalize the one-body reduced density matrix (OBDM) $\rho_1(\vb{r}, \vb{r}') = \langle {\psi}^\dagger(\vb{r}){\psi}(\vb{r}') \rangle = \sum_i \lambda_i \phi_i^*(\vb{r}) \phi_i(\vb{r}')$, with eigenvalues ordered as $\lambda_0\geq \lambda_1 \geq \dots$.
Defining the normalized occupations $p_i=\lambda_i/N$, with $\sum_i p_i=1$, we obtain the condensate wavefunction $\phi_0(\vb{r})$, the condensate depletion $1-p_0$, where $p_0=N_0/N$ is the condensate fraction, and the von Neumann and R\'enyi-2 single-particle entanglement entropies $S_1 = -\sum_i p_i \ln p_i$ and $S_2 = -\ln \sum_i p_i^2$, respectively.
In Fig.~\ref{fig: large N limit} (c-d), these are plotted versus $N$ (up to $N=1000$) along the same scaling trajectory.
For $N\le5$, they agree with the exact two-body and ED results, with mean deviations across couplings of at most $0.2\%$ for $N=2$ and $1.4\%$ for $3\le N\le5$.
As expected, all three decrease towards zero for large $N$, indicative of a pure condensate.

\begin{figure}[t]
    \includegraphics[width=\columnwidth]{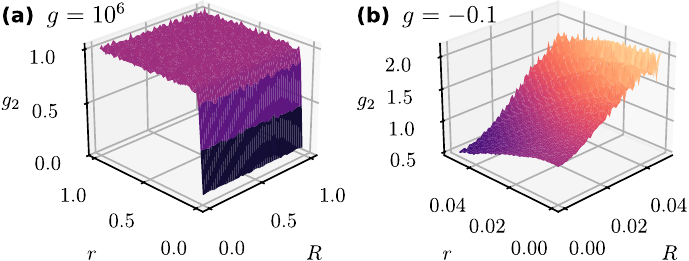}
    \caption{Ground-state two-body correlations $g_2(R,r)$ for $N=1000$ atoms in a 1D harmonic trap, with $R = |x_i + x_j|/2$ and $r = |x_i - x_j|$ the absolute centre-of-mass and relative coordinates of an atom pair, in units of $a_{\text{ho}}$.
    (a) Repulsion ($g=10^6$).
    (b) Attraction ($g=-0.1$).
    Surfaces are histogrammed from $N_{\rm MC}=10^6$ samples; the visible roughness is statistical noise.}
    \label{fig: g2}
\end{figure}

A cold Bose gas crosses over from a mean-field condensate to a fermionized (TG) gas as the repulsion grows.
Since experiments tune the coupling across this entire range \cite{Kinoshita2004, Haller2009}, a correct numerical description requires a method capable of interpolating accurately between both limits at large atom numbers.
Fig.~\ref{fig: large g limit} (a) shows the VMC energy per atom versus $g$ ($10^{-4} \leq g \leq 10^6$) for a repulsive 1D gas of $N=1000$ atoms, compared with GP, LL+LDA, and the TG energy $E_{\text{TG}} = N^2/2$.
The VMC energies lie within $1.8\%$ of the LL+LDA prediction across the entire range of $g$, with the largest deviation at $g\approx10$, where $\sigma_E$ [Fig.~\ref{fig: large g limit}(b)] is also large, and within $0.4\%$ of GP at weak interactions.
Whereas GP exceeds $E_{\text{TG}}/N$ from $g\approx 60$, the VMC energies always remain below it, saturating towards it at extremely strong interactions.
Fig.~\ref{fig: large g limit} (b) shows the standard deviation $\sigma_E$ of the sampled local energies.
By the zero-variance property of VMC, $\sigma_E^2=\langle E_\text{loc}^2\rangle-\langle E_\text{loc}\rangle^2$ vanishes if and only if the trial state is an exact eigenstate of $H$, making $\sigma_E$ an intrinsic accuracy measure of the ansatz.
It peaks at $g\sim10^2$ and decreases by several orders of magnitude towards both weak and strong interactions, confirming that the many-body trial state is exact in both limits.

Importantly, unlike GP or GP+LL, VMC also gives insights into the quantum correlations of the gas, quantified in Fig.~\ref{fig: large g limit} (c).
In addition to previously considered quantities, we plot the effective Schmidt rank $K = \left(\sum_i p_i^2\right)^{-1}$, which gives an estimate of the number of single-particle modes occupied in the many-body state \cite{RGrobe_1994}.
All these measures correctly characterize the crossover from a nearly-pure condensate at $g\approx 0$ to a fermionized gas, increasing monotonically with $g$ and saturating at large $g$.

\begin{figure}[t!]
    \includegraphics[width=\columnwidth]{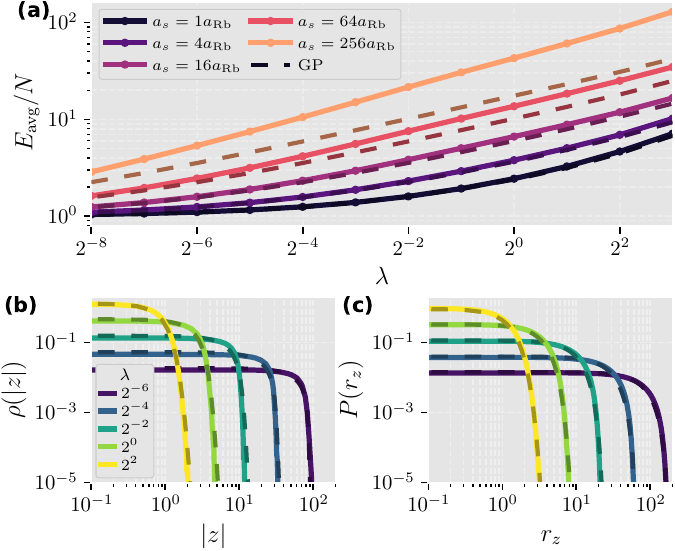}
    \caption{(a) Energy per atom of $N=1000$ hard-sphere atoms in anisotropic traps versus the anisotropy $\lambda=\omega_z/\omega_\perp$ for several $a_s$ (colors), compared with GP (darker dashed lines).
    (b) Axial distributions of atom--trap-centre distances $\rho(|z|)$ and (c) pair distances $P(r_z)$ for $a_s=16a_{\text{Rb}}$ and several $\lambda$, compared with GP.}
    \label{fig: aniso_lambda}
\end{figure}

To further quantify correlation effects, Fig.~\ref{fig: g2} shows the two-body correlation function $g_2(R,r)$ for a 1D gas of $N=1000$ atoms with repulsive and attractive $\delta$-interactions, in the absolute-valued centre-of-mass and relative coordinates $R$ and $r$ of an atom pair.
It is defined as the ratio of the two-body density to the product of one-body densities at positions $R\pm r/2$,
\begin{equation}
    g_2(R,r) = \frac{\langle \psi^\dagger(R+\frac{r}{2}) \psi^\dagger(R-\frac{r}{2})  \psi(R+\frac{r}{2}) \psi(R-\frac{r}{2}) \rangle}{\langle \psi^\dagger(R+\frac{r}{2}) \psi(R+\frac{r}{2}) \rangle \langle \psi^\dagger(R-\frac{r}{2}) \psi(R-\frac{r}{2}) \rangle}.
    \label{eq: g2}
\end{equation}
For the strongly repulsive gas with $g=10^6$ in Fig.~\ref{fig: g2} (a), effectively in the TG limit \cite{Girardeau1960}, $g_2(R,r)\to 0$ as $r \to 0$ across the plotted range of $R$, reflecting strong antibunching of the fermionized gas; the correlation hole recovers over the local interparticle spacing, $1/n(0) \simeq 0.07\,a_{\text{ho}}$ at the trap centre (consistent with exact TG results), beyond which $g_2$ saturates to unity.

For attractive interactions with $g=-0.1$ in Fig.~\ref{fig: g2} (b), $|g|N = 100$ places the gas deep in the self-bound regime, in which the atoms form a cluster of size $\sim 1/(|g|N) \sim 10^{-2}a_{\text{ho}}$, two orders of magnitude below the oscillator length and essentially unaffected by the trap, confining all structure in $g_2$ to $R, r \lesssim 0.04\,a_{\text{ho}}$.
Here $g_2$ exceeds unity and decreases with $r$, signalling the strong bunching expected under attraction.
Its pronounced growth with $R$ reflects the centre-of-mass delocalization of the cluster.
In the tails, the pair density remains finite as both atoms belong to the same cluster, while the product of one-body densities in the denominator of Eq.~\eqref{eq: g2} falls off faster.
The VMC energy per atom, $E/N\approx-413.7$, lies $0.7\%$ above the exact energy of the untrapped self-bound state, $E/N=-g^2(N^2-1)/24\approx-416.7$~\cite{McGuire1964}.

We now turn to the most general and experimentally ubiquitous case of a 3D anisotropic trap, whose aspect ratio is routinely tuned in experiments~\cite{Guo2024}.
An axially symmetric trap requires only three independent variational parameters, $\bm{\theta} = (\alpha, \beta_\bot, \beta_z)$.
Fig.~\ref{fig: aniso_lambda} (a) shows the energy per atom versus the trap anisotropy $\lambda = \omega_z/\omega_\bot$ (with $\omega_\bot=1$) for $a_s$ up to $256a_{\text{Rb}}$, compared with GP.
For an isotropic trap, the VMC energies again exceed GP, increasingly so with $a_s$, as expected from the LHY correction \cite{PhysRev.106.1135}.
As the trap is elongated and the gas density decreases, the VMC energies approach the mean-field GP predictions.
Conversely, as $\lambda$ increases and the trap shrinks, the deviation from GP grows with the gas density, as expected.
The axial distributions $\rho(|z|)$ and $P(r_z)$ for $a_s = 16a_{\text{Rb}}$ in Fig.~\ref{fig: aniso_lambda} (b-c) are in excellent agreement with GP for every anisotropy considered.
A single three-parameter ansatz thus describes the gas equally well in highly anisotropic geometries, from strongly elongated to oblate traps.

\section{Discussion}

We introduced a non-perturbative method for characterizing the quantum-correlated ground-state properties of ultracold bosons in experimentally relevant settings.
Our implementation was validated across 1D and 3D isotropic and anisotropic harmonic traps, over a wide range of repulsive interaction strengths and for a strongly attractive gas, and for atom numbers spanning several orders of magnitude, with reference to exact diagonalization, mean-field theory, and analytical results.
In 1D, where our ansatz remains exact in both the $g=0$ and $g=\infty$ limits, and while requiring only two independent variational parameters, we were able to study the crossover from an ideal condensate to a TG gas, quantifying the ground-state energy and quantum-correlation measures obtained from the single-particle density matrix for as many as $N=1000$ atoms at modest computational cost.

Given the simple Jastrow-type form of the variational ansatz, several natural extensions of our VMC algorithm are apparent.
These include augmenting the trial wavefunction with, e.g., species-dependent factors or Slater determinants, which would allow the simulation of Bose or Fermi mixtures.
Such extensions could enable studies of trimer and droplet formation in atomic and molecular gases \cite{Chuang2025, doi:10.1126/science.aao5686, PhysRevLett.134.093401, Zhang2026}.
Beyond VMC, the optimized trial state could also serve as a guiding function for diffusion Monte Carlo or as the projected state in path-integral ground-state (PIGS) methods~\cite{Sarsa2000}, where its proximity to the ground state could reduce, respectively, the mixed-estimate bias and the imaginary-time projection required.
Moreover, the exact solution underlying our ansatz describes the full two-body excitation spectrum, which may offer a route to excited-state properties of the gas at strong interactions~\cite{Haller2009}.
Combined with the scalability demonstrated here, such extensions would enable first-principles predictions for strongly correlated quantum gases at experimentally realistic conditions.

\begin{acknowledgments}
    This work is funded by the Engineering and Physical Sciences Research Council (Grant No.
    EP/R513143/1) and the European Union under Horizon Europe (project No. 101186579).
    Views and opinions expressed are however those of the author(s) only and do not necessarily reflect those of the European Union or the European Research Executive Agency.
    Neither the European Union nor the granting authority can be held responsible for them.
    K.R. acknowledges support of NCN (Polish National Science Center) grant 2024/53/B/ST2/02161.
    D.A.H. acknowledges the use of the UCL HPC facilities and associated support services in the completion of this work.
    Generative AI tools (Claude Opus 5.5, ChatGPT 6 Astra) were used to assist in code development (debugging, optimization, refactoring), figure production, and text editing (grammar, spelling, punctuation, stylistic improvement).
\end{acknowledgments}

\section*{Data availability}

The code that supports the findings of this article is publicly available at \url{https://github.com/dhryniuk/EPCAVMC}.
The simulation data are available from the authors upon reasonable request.

\bibliography{references.bib}

\end{document}


\renewcommand{\theequation}{S\arabic{equation}}
\renewcommand{\thefigure}{S\arabic{figure}}
\renewcommand{\thetable}{S\arabic{table}}

\title{Supplementary material for ``Scalable simulation of strongly interacting trapped ultracold bosons with an exact pair-correlated ansatz''}

\author{Dawid A. Hryniuk}
\email{d.hryniuk@ucl.ac.uk}
\affiliation{Department of Physics and Astronomy, University College London,
Gower Street, London, WC1E 6BT, United Kingdom}
\affiliation{Center for Theoretical Physics, Polish Academy of Sciences, Al. Lotników 32/46, 02-668 Warsaw, Poland}
\author{Mateusz Pańtak}
\affiliation{Center for Theoretical Physics, Polish Academy of Sciences, Al. Lotników 32/46, 02-668 Warsaw, Poland}
\author{Kazimierz Rzążewski}
\affiliation{Center for Theoretical Physics, Polish Academy of Sciences, Al. Lotników 32/46, 02-668 Warsaw, Poland}

\date{\today}

\maketitle

\supplementarytoc

\numbersupplementary

\subsection{Separation of the Hamiltonian}
\label{supp: separation}

In this section, we derive the centre-of-mass and relative Hamiltonians stated in Eqs. \eqref{M-eq: H_com} and \eqref{M-eq: H_rel} in the main text.
Let us define the map $A$ relating the absolute particle coordinates $\{\vb{r}_k\}_{k=1}^N$ and the pair-difference coordinates $\{\bm{r}_{p=(ij)} = \vb{r}_i - \vb{r}_j\}$ as $\bm{r}_{ij} = A_{ij,k}\vb{r}_k$, satisfying the properties
\begin{enumerate}
    \item $A_{ij,k} = \frac{\partial \bm{r}_{ij}}{\partial \vb{r}_k} = \delta_{ik} - \delta_{jk}$,
    \item $A^\top A = N \mathbbm{1}_N - \bm{1}\bm{1}^\top$,
    \item $(AA^\top)_{ij,kl} = \delta_{ik}-\delta_{il}-\delta_{jk}+\delta_{jl}$.
\end{enumerate}
Defining the $N\times d$ matrix $X_{i\alpha} = r_i^\alpha$ and the $\binom{N}{2}\times d$ matrix $R_{(ij)\alpha} = r_{ij}^\alpha$, where $R = AX$, one obtains for the quadratic terms proportional to the trap potential energy,
\begin{align}
    \sum_{i=1}^N \vb{r}_i^\top \Omega^2 \vb{r}_i &= \tr\left[X \Omega^2 X^\top\right] \\
    &= \frac{1}{N}\tr\left[X \Omega^2 X^\top (A^\top A + \bm{1}\bm{1}^\top)\right] \\
    &= \frac{1}{N}\tr\left[AX \Omega^2 (AX)^\top\right] + \frac{1}{N}\tr\left[\bm{1}^\top X \Omega^2 X^\top \bm{1}\right] \\
    &= \frac{1}{N}\tr\left[R \Omega^2 R^\top\right] + \frac{1}{N}(\bm{1}^\top X) \Omega^2 (X^\top \bm{1}) \\
    &= \frac{1}{N}\sum_{i<j}\bm{r}^\top_{ij} \Omega^2 \bm{r}_{ij} + N \vb{X}^\top \Omega^2 \vb{X},
\end{align}
where $X^\alpha = \frac{1}{N}\sum_{i=1}^N r_i^\alpha$ are the components of $\vb{X}$.
From the chain rule, one also obtains $\nabla_X = \frac{1}{N}\bm{1}\nabla^\top_{\vb{X}} + A^\top \nabla_R$.
Then, making use of $A\bm{1} = 0$ and $\bm{1}^\top\bm{1} = N$,
\begin{align}
    \sum_{i=1}^N \nabla_{\vb{r}_i}^2
    &= \tr\left[\nabla_X^\top \nabla_X\right] \\
    &= \tr\left[
    \left(\frac{1}{N}\nabla_{\vb{X}}\bm{1}^\top + \nabla_R^\top A\right)
    \left(\frac{1}{N}\bm{1}\nabla_{\vb{X}}^\top + A^\top \nabla_R\right)
    \right] \\
    &= \tr\left[
    \frac{1}{N^2}\nabla_{\vb{X}}\bm{1}^\top \bm{1}\nabla_{\vb{X}}^\top
    + \frac{1}{N}\nabla_{\vb{X}}\bm{1}^\top A^\top \nabla_R
    + \frac{1}{N}\nabla_R^\top A \bm{1}\nabla_{\vb{X}}^\top
    + \nabla_R^\top A A^\top \nabla_R
    \right] \\
    &= \tr\left[ \nabla_R^\top A A^\top \nabla_R \right] + \frac{1}{N}\tr\left[
    \nabla_{\vb{X}}\nabla_{\vb{X}}^\top\right] \\
    &= \sum_{i<j, k<l} (\delta_{ik}-\delta_{il}-\delta_{jk}+\delta_{jl}) \nabla_{\bm{r}_{ij}} \cdot \nabla_{\bm{r}_{kl}}- \frac{1}{\hbar^2 N}\vb{P}^2,
\end{align}
where we defined $\vb{P} = -i\hbar\sum_{i=1}^N \nabla_{\vb{r}_i}$.
Therefore, the Hamiltonian separates as $H = H_{\text{com}} + H_{\text{rel}}$ with
\begin{align}
    H_{\text{com}} &= \frac{1}{2Nm}\vb{P}^2 + \frac{Nm}{2} \vb{X}^\top \Omega^2 \vb{X}, \\
    H_{\text{rel}} &= -\frac{\hbar^2}{2m} \sum_{i<j, k<l} (\delta_{ik}-\delta_{il}-\delta_{jk}+\delta_{jl}) \nabla_{\bm{r}_{ij}} \cdot \nabla_{\bm{r}_{kl}} + \frac{m}{2N} \sum_{i<j}\bm{r}^\top_{ij} \Omega^2 \bm{r}_{ij} + \sum_{i<j} V(\bm{r}_{ij}).
\end{align}
In the above, $H_{\text{com}}$ is the Hamiltonian of a $d$-dimensional simple harmonic oscillator with mass $Nm$, with ground-state energy $E_{\text{com}} = \frac{\hbar}{2}\tr\Omega$ and wavefunction
\begin{equation}
    \Phi(\vb{X}) = \det\left(\frac{Nm\,\Omega}{\pi\hbar}\right)^{1/4}\exp\left(-\frac{Nm}{2\hbar}\vb{X}^\top\Omega\,\vb{X}\right).
\end{equation}

\subsection{Local kinetic energy estimator}
\label{supp: kinetic}

In this section, we derive the expressions for the local kinetic energy estimators in Eq. \eqref{M-eq: kinetic energy} in the main text.
From Eqs. \eqref{M-eq: local energy 2} and \eqref{M-eq: H_rel},
\begin{align}
    \frac{T \Psi_{\text{rel}}}{\Psi_{\text{rel}}}  &= -\frac{\hbar^2}{2m} \sum_{p,q} (AA^\top)_{pq} \frac{\nabla_{\bm{r}_{p}} \cdot \nabla_{\bm{r}_{q}} \Psi_{\text{rel}}}{\Psi_{\text{rel}}}\\
    &= -\frac{\hbar^2}{2m} \left[ \sum_{p} (AA^\top)_{pp} u_p + \sum_{p\neq q} (AA^\top)_{pq} \bm{v}_p \cdot \bm{v}_q \right]\\
    &= -\frac{\hbar^2}{2m}\sum_{p} (AA^\top)_{pp} \left[u_p - \|\bm{v}_p\|^2 \right] -\frac{\hbar^2}{2m} \sum_{p,q} (AA^\top)_{pq} \bm{v}_p \cdot \bm{v}_q\\
    &= -\frac{\hbar^2}{m}\sum_{p} \left[u_p - \|\bm{v}_p\|^2 \right] -\frac{\hbar^2}{2m} \|A^\top \bm{v}\|^2,
\end{align}
where we defined the Laplacian scalars $u_p = \nabla^2_{\bm{r}_{p}} f(\bm{r}_{p})/f(\bm{r}_{p})$ and gradient vectors $\bm{v}_p = \nabla_{\bm{r}_{p}}\ln f(\bm{r}_{p})$.
Here, $\bm{v}$ collects the pair-gradient vectors, and $\|A^\top\bm{v}\|$ denotes the Frobenius norm over particle and spatial indices.
The computation of the local kinetic energy estimator therefore requires the computation of $O(N^2)$ first- or second-order derivatives per sample (sum over $p$), and memorization of the $N\times d$ array $A^\top \bm{v}$, whose entries can be efficiently updated also in $O(N^2)$ time alongside computing $u_p$ and $\bm{v}_p$.
We now find the explicit expressions for $u_p$ and $\bm{v}_p$ for the three trapping potentials considered in this work.

\subsubsection{1D trap}

Let us begin by considering the simplest 1D case, with $\tilde{r}_p^2 = r_p^2/a_0^2$ and $z_p = \alpha^2\tilde{r}_p^2$.
Recall
\begin{equation}
    f_p = U\left(-\nu, \frac{1}{2}, \alpha^2 \tilde{r}_p^2\right)
    \exp\left(-\frac{\beta}{2}\tilde{r}_p^2\right).
\end{equation}
For brevity, denote $U_n = U\left(-\nu+n, \frac{1}{2}+n, z_p\right)$.
From the product and chain differentiation rules, one easily obtains
\begin{align}
    v_p &= \frac{\partial_{r_p} f_p}{f_p} \\
    &= \left(\frac{2\alpha^2\nu U_1}{U_0} - \beta\right)\frac{r_p}{a_0^2} \\
    &\defeq \Lambda_p \frac{r_p}{a_0^2}.
\end{align}
Similarly,
\begin{align}
    u_p &= \frac{\partial^2_{r_p} f_p}{f_p} \\
    &= v_p^2 + \partial_{r_p} v_p \\
    &= v_p^2 + \frac{\partial \Lambda_p}{\partial r_p}\frac{r_p}{a_0^2}
    + \frac{\Lambda_p}{a_0^2}.
\end{align}
We evaluate $\partial\Lambda_p/\partial r_p$ using the quotient rule:
\begin{align}
    \frac{\partial\Lambda_p}{\partial r_p}
    &= 2\alpha^2\nu\frac{\partial}{\partial r_p}\left(\frac{U_1}{U_0}\right) \\
    &= 2\alpha^2\nu\frac{U_0\,\partial_{r_p}U_1 - U_1\,\partial_{r_p}U_0}{U_0^2} \\
    &= 2\alpha^2\nu\frac{U_0(\nu-1)U_2 - U_1\cdot\nu U_1}{U_0^2}
    \cdot\frac{2\alpha^2 r_p}{a_0^2}
\end{align}
where we used $\partial_{z_p} U_n = (\nu-n)U_{n+1}$ and $\partial z_p/\partial r_p = 2\alpha^2 r_p/a_0^2$.
We eliminate $U_2$ using the Tricomi ODE $z_pU_0'' + (\frac{1}{2} - z_p)U_0' + \nu U_0 = 0$, which gives
\begin{equation}
    (\nu-1)U_2 = \frac{\left(z_p - \frac{1}{2}\right)U_1 - U_0}{z_p}.
\end{equation}
Substituting,
\begin{equation}
    \frac{\partial\Lambda_p}{\partial r_p} =
    \frac{4\alpha^4\nu r_p}{a_0^2}
    \left[\frac{(z_p-\frac{1}{2})U_1/U_0 - 1}{z_p}
    - \nu\frac{U_1^2}{U_0^2}\right].
\end{equation}
Inserting into $u_p$, using $4\alpha^4 r_p^2/(a_0^4 z_p) = 4\alpha^2/a_0^2$, and expanding $\Lambda_p^2 - 4\alpha^4\nu^2 U_1^2/U_0^2 = -4\alpha^2\nu\beta U_1/U_0 + \beta^2$:
\begin{align}
    u_p &= v_p^2 + \frac{\partial\Lambda_p}{\partial r_p}\frac{r_p}{a_0^2}
    + \frac{\Lambda_p}{a_0^2} \\
    &= \frac{r_p^2}{a_0^4}
    \left[\frac{4\alpha^2\nu}{\tilde{r}_p^2}
    \left(\left(z_p - \tfrac{1}{2}\right)\frac{U_1}{U_0} - 1\right)
    - \frac{4\alpha^2\nu\beta U_1}{U_0} + \beta^2\right] + \frac{\Lambda_p}{a_0^2}.
\end{align}

\subsubsection{3D isotropic trap}
\label{supp: kinetic 3d iso}

Two changes to the previous results occur in the 3D isotropic case.
First, the second arguments of the confluent hypergeometric functions $U_n$ increase by 1, so now $U_n = U\left(-\nu+n, \frac{3}{2}+n, z_p\right)$, with $z_p=\alpha^2\tilde{r}_p^2$.
Second, the Laplacian in spherical coordinates becomes
\begin{align}
    u_p &= \frac{f''(r_{p})}{f(r_{p})} + \frac{2}{r_p} \frac{f'(r_{p})}{f(r_{p})} \\
    &= \left(\frac{f'(r_p)}{f(r_p)}\right)^2
    + \frac{\partial \Lambda_p}{\partial r_p}\frac{r_p}{a_0^2}
    + \frac{\Lambda_p}{a_0^2} + \frac{2\Lambda_p}{a_0^2} \\
    &= \|\bm{v}_p\|^2 + \frac{\partial \Lambda_p}{\partial r_p}\frac{r_p}{a_0^2}
    + \frac{3\Lambda_p}{a_0^2},
\end{align}
where $\bm{v}_p = \Lambda_p\bm{r}_p/a_0^2$, with $\Lambda_p$ now a function of the new $U_n$.
The computation of $\partial\Lambda_p/\partial r_p$ proceeds identically, but the Tricomi ODE now reads $z_pU_0'' + (\frac{3}{2} - z_p)U_0' + \nu U_0 = 0$, giving
\begin{equation}
    (\nu-1)U_2 = \frac{\left(z_p - \frac{3}{2}\right)U_1 - U_0}{z_p},
\end{equation}
and hence, eliminating $U_2$,
\begin{equation}
    \frac{\partial\Lambda_p}{\partial r_p} =
    \frac{4\alpha^4\nu r_p}{a_0^2}
    \left[\frac{(z_p-\frac{3}{2})U_1/U_0 - 1}{z_p}
    - \nu\frac{U_1^2}{U_0^2}\right].
\end{equation}
Inserting into $u_p$, using $4\alpha^4 r_p^2/(a_0^4 z_p) = 4\alpha^2/a_0^2$, and expanding $\Lambda_p^2 - 4\alpha^4\nu^2 U_1^2/U_0^2 = -4\alpha^2\nu\beta U_1/U_0 + \beta^2$:
\begin{align}
    u_p &= \|\bm{v}_p\|^2 + \frac{\partial\Lambda_p}{\partial r_p}\frac{r_p}{a_0^2}
    + \frac{3\Lambda_p}{a_0^2} \\
    &= \frac{r_p^2}{a_0^4}
    \left[\frac{4\alpha^2\nu}{\tilde{r}_p^2}
    \left(\left(z_p - \tfrac{3}{2}\right)\frac{U_1}{U_0} - 1\right)
    - \frac{4\alpha^2\nu\beta U_1}{U_0} + \beta^2\right] + \frac{3\Lambda_p}{a_0^2}.
\end{align}
The only differences from the 1D result are the replacement $\tfrac{1}{2}\to\tfrac{3}{2}$ (in the Tricomi ODE and $U_n$ parameters) and the coefficient $1/a_0^2 \to 3/a_0^2$ in the $\Lambda_p$ term, which reflects the three degrees of freedom contributing to the Laplacian.

\subsubsection{3D anisotropic trap}

In the 3D anisotropic case, we work in Cartesian coordinates with the pair factor of Eq.~\eqref{M-eq: pair product anisotropic},
\begin{equation}
    f_p = U\!\left(-\nu,\tfrac{3}{2},z_p\right)\exp\!\left(-\frac{1}{2}\sum_n \beta_n \tilde{r}_{pn}^2\right),
    \qquad z_p = \alpha^2\tilde{r}_p^2,
\end{equation}
where $\tilde{r}_p = r_p/a_0$ with $a_0^2=\frac{1}{3}\sum_n a_n^2$, $\tilde{r}_{pn} = r_{pn}/a_n$, and $U_n(z_p) = U(-\nu+n,\frac{3}{2}+n,z_p)$.
Since only the Gaussian factor depends on direction, the gradient components separate into a radial Tricomi part and a direction-dependent Gaussian part,
\begin{equation}
    v_{pn} = 2\alpha^2\nu\frac{U_1(z_p)}{U_0(z_p)}\frac{r_{pn}}{a_0^2} - \beta_n\frac{r_{pn}}{a_n^2}.
\end{equation}
For the Laplacian contribution $u_p = \sum_n \partial^2_{r_{pn}}f_p/f_p = \sum_n\left[v_{pn}^2 + \partial_{r_{pn}}v_{pn}\right]$, the derivative of the Tricomi part follows as in Sec.~\ref{supp: kinetic 3d iso}, with $\partial z_p/\partial r_{pn} = 2\alpha^2 r_{pn}/a_0^2$:
\begin{equation}
    \frac{\partial}{\partial r_{pn}}\left[2\alpha^2\nu\frac{U_1(z_p)}{U_0(z_p)}\right] = \frac{4\alpha^4\nu r_{pn}}{a_0^2}\left[\frac{\left(z_p-\frac{3}{2}\right)U_1(z_p)/U_0(z_p) - 1}{z_p} - \nu\frac{U_1^2(z_p)}{U_0^2(z_p)}\right].
\end{equation}
Inserting this into $u_p$, the Tricomi contribution $4\alpha^4\nu^2 r_p^2U_1^2(z_p)/[a_0^4U_0^2(z_p)]$ to $\sum_n v_{pn}^2$ cancels the $-\nu U_1^2(z_p)/U_0^2(z_p)$ term, and using $4\alpha^4 r_p^2/(a_0^4 z_p) = 4\alpha^2/a_0^2$ we obtain
\begin{equation}
    u_p = \frac{4\alpha^2\nu}{a_0^2}\left[\left(z_p-\tfrac{3}{2}\right)\frac{U_1(z_p)}{U_0(z_p)} - 1\right] + \frac{6\alpha^2\nu}{a_0^2}\frac{U_1(z_p)}{U_0(z_p)} + \sum_n\left[\frac{\beta_n^2 r_{pn}^2}{a_n^4} - \frac{4\alpha^2\nu\beta_n r_{pn}^2}{a_0^2a_n^2}\frac{U_1(z_p)}{U_0(z_p)} - \frac{\beta_n}{a_n^2}\right].
\end{equation}
In the isotropic limit $a_n \to a_0$, $\beta_n \to \beta$, this reduces to the radial expression of Sec.~\ref{supp: kinetic 3d iso}.

\subsection{Metropolis Markov-chain Monte Carlo}
\label{supp: metropolis}

We sample new configurations using a simple Markov-chain Monte Carlo protocol with local single-atom Metropolis updates, which we found to sample the Hilbert space more efficiently than global all-atom updates.
Specifically, the position of one randomly chosen particle $\vb{r}_k\in\vb{R}$ is shifted to a neighbouring position $\vb{r}_k'\in\vb{R}'$, drawn from a uniform probability density centered on $\vb{r}_k$.
The proposed move is accepted with probability $\min\{1,|\Psi(\vb{R}'; \bm{\theta})|^2/|\Psi(\vb{R}; \bm{\theta})|^2\}$.
The probability ratio is
\begin{align}
    \frac{|\Psi(\vb{R}'; \bm{\theta})|^2}{|\Psi(\vb{R}; \bm{\theta})|^2} &= \frac{|\Phi(\vb{X}')|^2}{|\Phi(\vb{X})|^2} \prod_{i < j} \frac{|f(\bm{r}_{ij}'; \bm{\theta})|^2}{|f(\bm{r}_{ij}; \bm{\theta})|^2} \\
    &= \frac{|\Phi(\vb{X}')|^2}{|\Phi(\vb{X})|^2} \prod_{i < k} \frac{|f(\vb{r}_i - \vb{r}_k'; \bm{\theta})|^2}{|f(\vb{r}_i - \vb{r}_k; \bm{\theta})|^2} \prod_{k < j} \frac{|f(\vb{r}_k' - \vb{r}_j; \bm{\theta})|^2}{|f(\vb{r}_k - \vb{r}_j; \bm{\theta})|^2}.
\end{align}
All ratios of pair factors not involving particle $k$ cancel.
Therefore, the acceptance probability can be computed in $O(N)$ time.
Each iteration, the mean acceptance rate is recorded and the width of the uniform proposal density adjusted to maintain an acceptance rate of around $30\%$.

A large number of local one-atom updates are performed to generate samples with low autocorrelation, measured by the autocorrelation time
\begin{equation}
    \tau = 1 + 2\sum_{t=1}^W \frac{C(t)}{C(0)},
\end{equation}
where $M$ denotes the number of samples, $W<M$ is the autocorrelation truncation window, and the sample autocorrelation function $C(t)$ is defined as
\begin{equation}
    C(t) = \frac{1}{M-t}\sum_{i=1}^{M-t} (E_i - \langle E\rangle)(E_{i+t} - \langle E\rangle),
\end{equation}
where $E_i$ is the local energy of the $i^\text{th}$ sample, and $\langle E\rangle$ is the sample mean.
The Monte Carlo standard errors are adjusted for $\tau>1$.
To reduce the computational cost, the gradient contributions are computed only every $\kappa$ sample, where we set $\kappa_i \leftarrow \kappa_{i-1} \times \text{clamp}(1+\tau-\tau_0, 0.8, 2.0)$ per iteration, up to some maximum allowed value, usually set to $\kappa = 10^4$.
Here, $\tau_0$ is the target autocorrelation time, usually set at $\tau_0 = 1.2$.
This procedure generally reduces the simulation times as the autocorrelations times decrease when approaching the energy minimum, while still ensuring that the gradient contributions are computed on effectively independent samples.

Additionally, during optimization, the final samples from a previous iteration are reused as the starting conditions for sampling in the subsequent iteration (after undergoing burn-in with the now updated variational parameters), to further reduce total computational cost.

\subsection{Gradients}
\label{supp: gradients}

In this section, we derive the per-pair log-derivative contributions $\Delta_{\theta_i}(\tilde{\bm{r}}_p) = \partial_{\theta_i}\ln f(\bm{r}_p;\bm{\theta})$ used in the optimization of the variational parameters, such that the full log derivative is $\partial_{\theta_i}\ln\Psi(\vb{R};\bm{\theta})=\sum_p\Delta_{\theta_i}(\tilde{\bm{r}}_p)$.

\subsubsection{1D trap}

The per-pair $\beta$ log-derivative is straightforward:
\begin{align}
    \Delta_\beta(\tilde{r}_p)\;
    &\defeq \partial_\beta \ln f(r_p;\alpha,\beta) \\
    &= -\frac{1}{2}\tilde{r}_p^2.
\end{align}

The expression for the $\alpha$ log-derivative is more involved owing to the argument $-\nu_\alpha$ in the functions $U$ and the transcendental equation it is a root of.
Let $z_p=\alpha^2\tilde{r}_p^2$ and $U_n(z_p)=U(-\nu+n,\frac{1}{2}+n,z_p)$, and define the function
\begin{equation}
    F(\alpha,\nu) = \alpha \Gamma(-\nu+\tfrac{1}{2}) + \frac{g}{2\hbar\omega a_0}\Gamma(-\nu),
\end{equation}
where $F(\alpha,\nu_\alpha) = 0$ imposes the Lieb-Liniger condition.
Then, the log-derivative can be expressed as
\begin{align}
    \Delta_\alpha(\tilde{r}_p)\;
    &\defeq \partial_\alpha \ln f(r_p;\alpha,\beta) \\
    &= \partial_\alpha \ln U(-\nu,\tfrac{1}{2},z_p) \\
    &= \left[
    \frac{\partial\nu}{\partial\alpha}\frac{\partial}{\partial\nu}
    +2\alpha\tilde{r}_p^2\frac{\partial}{\partial z_p}
    \right]
    \ln U(-\nu,\tfrac{1}{2},z_p),
\end{align}
where, from the implicit function theorem, it follows that
\begin{align}
    \frac{\partial \nu}{\partial \alpha} &= -\frac{\partial_\alpha F}{\partial_\nu F} \\
    &= -\frac{\Gamma\left(-\nu+\tfrac{1}{2}\right)}{\partial_\nu\left[\alpha\,\Gamma\left(-\nu+\tfrac{1}{2}\right) + \dfrac{g}{2\hbar\omega a_0}\,\Gamma(-\nu)\right]} \\
    &= -\frac{\Gamma\left(-\nu+\tfrac{1}{2}\right)}{-\alpha\,\psi\left(-\nu+\tfrac{1}{2}\right)\Gamma\left(-\nu+\tfrac{1}{2}\right) - \dfrac{g}{2\hbar\omega a_0}\,\psi(-\nu)\,\Gamma(-\nu)},
\end{align}
where we used $\partial_\nu\,\Gamma(-\nu+c) = -\psi(-\nu+c)\,\Gamma(-\nu+c)$, where $\psi$ is the digamma function.
Substituting $\tfrac{g}{2\hbar\omega a_0}\,\Gamma(-\nu) = -\alpha\,\Gamma\left(-\nu+\tfrac{1}{2}\right)$, which follows from $F(\alpha,\nu)=0$, and cancelling the common term $\Gamma\left(-\nu+\tfrac{1}{2}\right)$, one obtains
\begin{align}
    \frac{\partial \nu}{\partial \alpha}
    &= -\frac{1}{\alpha\left[\psi(-\nu)-\psi\left(-\nu+\tfrac{1}{2}\right)\right]} \\
    &= \frac{1}{\alpha}\left[\psi\left(-\nu+\tfrac{1}{2}\right) - \psi(-\nu)\right]^{-1}.
\end{align}
Therefore, one finds,
\begin{align}
    \Delta_\alpha(\tilde{r}_p) = \frac{1}{\alpha}\left[\psi\left(-\nu+\tfrac{1}{2}\right) - \psi(-\nu)\right]^{-1} \frac{\partial}{\partial \nu} \ln U(-\nu, \tfrac{1}{2}, z_p) + 2\nu\alpha \tilde{r}_p^2 \frac{U_1(z_p)}{U_0(z_p)}.
\end{align}
No simple closed-form expression for $\partial_\nu \ln U(-\nu, \tfrac{1}{2}, z_p)$ is known; instead, we evaluate it numerically via the complex-step trick, i.e.
\begin{equation}
    \partial_\nu \ln U(-\nu, \tfrac{1}{2}, z_p) = -\frac{\Im U(-\nu + i\varepsilon, \tfrac{1}{2}, z_p)}{\varepsilon U(-\nu, \tfrac{1}{2}, z_p)} + O(\varepsilon^2)
\end{equation}
for some small $\varepsilon \sim 10^{-8}$.
We denote $U_{0+i\varepsilon}(z_p) := \varepsilon^{-1}\Im U(-\nu + i\varepsilon, \tfrac{1}{2}, z_p)$ for brevity.
Therefore, the final expression for the $\alpha$ log-derivative is
\begin{equation}
    \Delta_\alpha(\tilde{r}_p) = -\frac{1}{\alpha}\left[\psi\left(-\nu+\tfrac{1}{2}\right) - \psi(-\nu)\right]^{-1} \frac{U_{0+i\varepsilon}(z_p)}{U_0(z_p)} + 2\nu\alpha \tilde{r}_p^2 \frac{U_1(z_p)}{U_0(z_p)}.
\end{equation}

\subsubsection{3D isotropic trap}
\label{supp: gradients 3d iso}

In a 3D isotropic trap, let $z_p=\alpha^2\tilde{r}_p^2$, $U_n(z)=U(-\nu+n,\frac{3}{2}+n,z)$, and $U_{0+i\varepsilon}(z)=\varepsilon^{-1}\Im U(-\nu+i\varepsilon,\frac{3}{2},z)$.
The per-pair $\beta$ log-derivative remains
\begin{equation}
    \Delta_\beta(\tilde{\bm{r}}_p)
    := \partial_\beta \ln f(\bm{r}_p;\alpha,\beta)
    = -\frac{1}{2}\tilde{r}_p^2.
\end{equation}
The $\alpha$ log-derivative is modified by the hard-sphere contact condition.
We define
\begin{equation}
    F(\alpha,\nu) = U\left(-\nu, \frac{3}{2}, z_s\right),
    \qquad z_s \equiv \alpha^2 \frac{a_s^2}{a_0^2},
\end{equation}
again imposing $F(\alpha,\nu) = 0$.
Then, applying the implicit function theorem,
\begin{align}
    \frac{\partial \nu}{\partial \alpha} &= -\frac{\partial_\alpha F}{\partial_\nu F} \\
    &= -\frac{2\nu\alpha a_s^2}{a_0^2}\frac{U(-\nu+1, \frac{5}{2}, z_s)}{\partial_\nu U(-\nu, \frac{3}{2}, z_s)}.
\end{align}
Therefore,
\begin{align}
    \Delta_\alpha(\tilde{\bm{r}}_p)\; &\defeq \partial_\alpha \ln f(\bm{r}_p;\alpha,\beta) \\
    &= \left[\frac{\partial \nu}{\partial \alpha}\frac{\partial}{\partial \nu} + 2\alpha \tilde{r}_p^2\frac{\partial}{\partial z_p}\right] \ln U(-\nu, \tfrac{3}{2}, z_p) \\
    &= -\frac{2\nu\alpha a_s^2}{a_0^2}\frac{U(-\nu+1, \frac{5}{2}, z_s)}{\partial_\nu U(-\nu, \frac{3}{2}, z_s)} \partial_\nu \ln U(-\nu, \tfrac{3}{2}, z_p) + 2\nu\alpha \tilde{r}_p^2 \frac{U_1(z_p)}{U_0(z_p)} \\
    &= -\frac{2\nu\alpha a_s^2}{a_0^2}\frac{U_1(z_s)}{U_{0+i\varepsilon}(z_s)} \frac{U_{0+i\varepsilon}(z_p)}{U_0(z_p)} + 2\nu\alpha \tilde{r}_p^2 \frac{U_1(z_p)}{U_0(z_p)}.
\end{align}

\subsubsection{3D anisotropic trap}

In the anisotropic case, only the Gaussian parameters are direction dependent ($\beta \to \beta_n$, $n \in \{x,y,z\}$), while $\alpha$ remains a single parameter.
The $\beta_n$ log-derivatives are
\begin{equation}
    \Delta_{\beta_n}(\tilde{\bm{r}}_p)
    := \partial_{\beta_n} \ln f_p
    = -\frac{1}{2}\tilde{r}_{pn}^2,
\end{equation}
with $\tilde{r}_{pn} = r_{pn}/a_n$.
For the axially symmetric traps of the main text, $\beta_x = \beta_y \equiv \beta_\perp$, so that $\Delta_{\beta_\perp} = \Delta_{\beta_x} + \Delta_{\beta_y}$.
The Tricomi factor depends on the pair separation only through $z_p = \alpha^2\tilde{r}_p^2$, with $\tilde{r}_p = r_p/a_0$, and $\nu$ is fixed by the same spherical hard-sphere condition $U(-\nu,\frac{3}{2},z_s) = 0$, with $z_s = \alpha^2 a_s^2/a_0^2$, as in Eq.~\eqref{M-eq: hard sphere condition}.
The $\alpha$ log-derivative $\Delta_\alpha(\tilde{\bm{r}}_p)$ is therefore identical to that of the 3D isotropic case derived in Sec.~\ref{supp: gradients 3d iso}.

\subsection{Interpolants}
\label{supp: interpolants}

Cumulatively, a single optimization iteration involves the repeated computation of the three confluent hypergeometric functions $U_0$, $U_1$ and $U_{0+i\varepsilon}$.
This becomes expensive as the number of atoms and Monte Carlo samples increases.
To reduce the overall computational cost, we make use of interpolants for the $U_0$, $U_1$ and $U_{0+i\varepsilon}$ functions.
An optimal approach was found to involve constructing the interpolants at the beginning of each iteration over a grid in the scalar argument $\alpha\tilde{r}$ at constant grid spacing density, bounded by a small cutoff on the order of $10^{-5}\sim 10^{-3}$ for small $\alpha\tilde{r}$ (necessitated by the divergence of the confluent hypergeometric functions for very small arguments) and by a suitably large cutoff for large $\alpha\tilde{r}$ determined by the maximum recorded atom separation from a previous set of Monte Carlo samples.

\begin{figure}[!ht]
    \centering
    \includegraphics[width=0.9\textwidth]{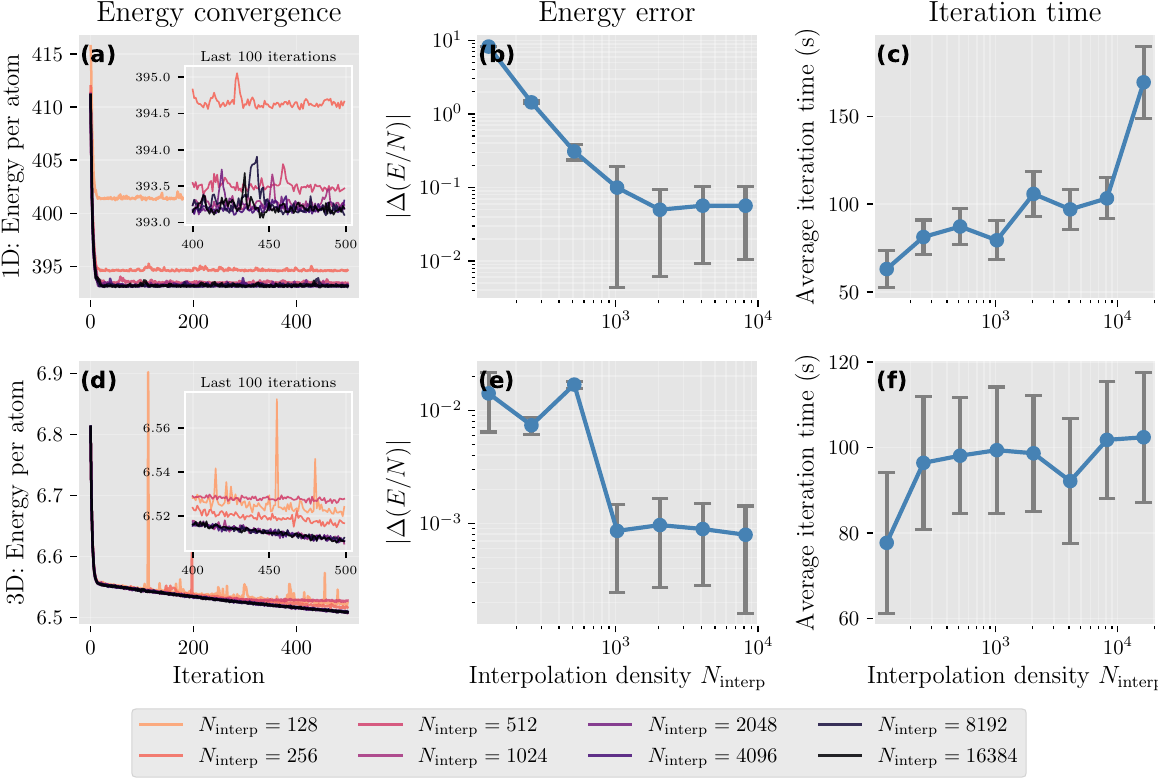}
    \caption{Convergence with the interpolation density $N_{\rm interp}$ (grid points per unit $\alpha\tilde{r}$) for a 1D gas with $g=100$ (top row) and a 3D gas of hard spheres with $a_s=16a_{\text{Rb}}$ (bottom row).
    All runs use $N=1000$ atoms, 10 MPI processes and $2500$ Monte Carlo samples per iteration ($250$ per process).
    (a,d) Variational energy per atom over 500 stochastic-reconfiguration iterations for $N_{\rm interp}=128$--$16384$; the insets show the final 100 iterations.
    (b,e) Absolute difference between the energy per atom at each $N_{\rm interp}$ and that obtained at the same iteration with $N_{\rm interp}=16384$, averaged over the final 50 iterations; error bars show the corresponding standard deviation.
    (c,f) Average wall-clock time per iteration as a function of $N_{\rm interp}$.
    For $N_{\rm interp}\gtrsim 10^3$, the energy differences are comparable to their iteration-to-iteration fluctuations, while the iteration time depends only weakly on the interpolation density over this range.}
    \label{fig: interpolant convergence}
\end{figure}

In Fig. \ref{fig: interpolant convergence} we study the convergence of the variational ground-state energy with the interpolation density.
The agreement with the densest $N_{\rm interp}=16384$ results improves as $N_{\rm interp}$ is increased, before appearing to plateau at around $N_{\rm interp}\sim 1000$.
Increasing the interpolation density leads to a slight corresponding increase in the iteration time in 1D, whereas the additional cost appears negligible in 3D.
We therefore find that a density of approximately 1000 points per unit $\alpha\tilde{r}$ is sufficient to converge the energy to within the statistical error bars for the results presented in the main text.

\subsection{Exact diagonalization in the harmonic-oscillator basis}
\label{supp: ED}

This section describes the exact-diagonalization (ED) calculations that provide few-body benchmarks for the one-dimensional system of the main text.
By ED we mean that, at a fixed harmonic-oscillator (HO) cutoff, the lowest eigenpair of the Hamiltonian projected onto the truncated bosonic Hilbert space is computed numerically, with the error due to the finite basis controlled by cutoff-convergence analysis and by extrapolation to the infinite-basis limit.

In one dimension and HO units ($\hbar=m=\omega=1$), the Hamiltonian of the main text reads
\begin{equation}
    H=\sum_{\alpha=1}^{N}h_\alpha
    +g\sum_{1\le\alpha<\beta\le N}V_{\alpha\beta},
    \qquad
    h_\alpha=-\frac12\frac{\partial^2}{\partial x_\alpha^2}+\frac12x_\alpha^2,
    \qquad
    V_{\alpha\beta}=\delta(x_\alpha-x_\beta),
    \label{eq:ED-H}
\end{equation}
so $g$ is the strength of the pairwise contact interaction in oscillator units.

\subsubsection{Basis and bosonic symmetry}
\label{app:ED-basis}

The one-particle HO orbitals
\begin{equation}
    \varphi_n(x)=\pazocal N_n H_n(x)\,e^{-x^2/2},
    \qquad
    \pazocal N_n=\bigl(\sqrt{\pi}\,2^n n!\bigr)^{-1/2},
    \label{eq:ED-orbitals}
\end{equation}
where $H_n$ are Hermite polynomials, satisfy $h\,\varphi_n=(n+\tfrac12)\varphi_n$.
We retain $n=0,1,\dots,n_{\max}$, i.e., $M=n_{\max}+1$ orbitals per particle.
A product state $|\mathbf j\rangle=|j_1,\dots,j_N\rangle$ has the wave function $\prod_{\alpha}\varphi_{j_\alpha}(x_\alpha)$, while a general state of the truncated product space has the wave function $\sum_{\mathbf j}T_{\mathbf j}\prod_{\alpha}\varphi_{j_\alpha}(x_\alpha)$, whose coefficients form a tensor $T_{\mathbf j}=T_{j_1\cdots j_N}$ with $M^N$ entries.

For bosons only permutation-symmetric tensors are physical, and we store one coefficient per ordered index tuple
\begin{equation}
    \mathbf i=(i_1,\dots,i_N),\qquad 0\le i_1\le i_2\le\dots\le i_N\le n_{\max}.
\end{equation}
Let $\nu_n(\mathbf i)$ be the number of entries of $\mathbf i$ equal to $n$, and let $\operatorname{Perm}(\mathbf i)$ be the set of distinct tuples obtained by permuting the entries of $\mathbf i$.
The number of elements of this set is
\begin{equation}
    \mu_{\mathbf i}=\bigl|\operatorname{Perm}(\mathbf i)\bigr|
    =\frac{N!}{\prod_n\nu_n(\mathbf i)!}.
\end{equation}
The normalized bosonic basis states
\begin{equation}
    |\mathbf i\rangle_B=\mu_{\mathbf i}^{-1/2}
    \sum_{\mathbf j\in\operatorname{Perm}(\mathbf i)}|\mathbf j\rangle
    \label{eq:ED-bosonic-basis}
\end{equation}
are the Fock states with orbital occupations $\nu_n(\mathbf i)$.
A state is stored as the coefficient vector $\mathbf c$ in
\begin{equation}
    |\Psi\rangle=\sum_{\mathbf i}c_{\mathbf i}\,|\mathbf i\rangle_B,
    \qquad
    \dim\pazocal H_B=\binom{M+N-1}{N}=\binom{n_{\max}+N}{N}.
    \label{eq:ED-dim}
\end{equation}
The combinatorial growth of $\dim\pazocal H_B$ limits the attainable cutoffs (Table~\ref{tab:ED-params}).

Since every $\mathbf j$ belongs to $\operatorname{Perm}(\mathbf i)$ for exactly one ordered tuple $\mathbf i$, the two representations are related by the map $U$ (``scatter''), which expands $\mathbf c$ into the symmetric tensor of the same state, and by its transpose $U^{\mathsf T}$ (``gather''):
\begin{equation}
    (U\mathbf c)_{\mathbf j}=\frac{c_{\mathbf i}}{\sqrt{\mu_{\mathbf i}}}
    \quad\text{for }\mathbf j\in\operatorname{Perm}(\mathbf i),
    \qquad
    (U^{\mathsf T}T)_{\mathbf i}=\frac{1}{\sqrt{\mu_{\mathbf i}}}
    \sum_{\mathbf j\in\operatorname{Perm}(\mathbf i)}T_{\mathbf j}.
    \label{eq:ED-scatter-gather}
\end{equation}
Here $U^{\mathsf T}U=\mathbb{I}$ and $UU^{\mathsf T}$ is the projector onto symmetric tensors, so $U$ and $U^{\mathsf T}$ are mutually inverse between coefficient vectors and symmetric tensors; for symmetric $T$, $(U^{\mathsf T}T)_{\mathbf i}=\sqrt{\mu_{\mathbf i}}\,T_{\mathbf i}$.

\subsubsection{Matrix elements of the contact interaction}
\label{app:ED-G}

In the orbital basis, the contact interaction has the two-particle matrix elements
\begin{equation}
    \langle i,j|V_{12}|k,l\rangle=G_{ijkl}
    \equiv\int_{-\infty}^{\infty}\!dx\,
    \varphi_i(x)\varphi_j(x)\varphi_k(x)\varphi_l(x).
    \label{eq:ED-G}
\end{equation}
The rank-four tensor $G$, symmetric in all four indices, is thus the representation of the contact interaction in the truncated basis and involves no additional approximation.

By Eq.~\eqref{eq:ED-orbitals}, the integrand is $e^{-2x^2}$ times a polynomial of degree $i+j+k+l\le4n_{\max}$.
Writing $\varphi_n(x)=e^{-x^2/2}\chi_n(x)$ with $\chi_n=\pazocal N_nH_n$, substituting $y=\sqrt2\,x$, and using Gauss--Hermite nodes $y_q$ and weights $w_q$ (weight function $e^{-y^2}$, $q=1,\dots,Q$), we obtain
\begin{equation}
    G_{ijkl}=\sum_{q=1}^{Q}\frac{w_q}{\sqrt2}\,
    \chi_i(x_q)\,\chi_j(x_q)\,\chi_k(x_q)\,\chi_l(x_q),
    \qquad x_q=\frac{y_q}{\sqrt2}.
    \label{eq:ED-GH}
\end{equation}
A $Q$-point rule integrates polynomials of degree up to $2Q-1$ exactly, so Eq.~\eqref{eq:ED-GH} is exact in exact arithmetic for $Q\ge2n_{\max}+1$.
We use $Q=\max(2n_{\max}+12,\,128)$, which exceeds this requirement by a safety margin.

In practice the index pairs are combined into composite indices: with $F_{q,(ij)}=\chi_i(x_q)\chi_j(x_q)$ and $W=\mathrm{diag}(w_q/\sqrt2)$, the $M^2\times M^2$ pair matrix is $G=F^{\mathsf T}WF$, which is then replaced by $(G+G^{\mathsf T})/2$ to remove round-off asymmetry.
Since $G$ depends only on $n_{\max}$, it is computed once per basis and reused for all couplings and all matrix--vector products.

\subsubsection{Matrix-free Lanczos diagonalization}
\label{app:ED-lanczos}

In the product-index representation, the one-body part of $H$ is diagonal, and the contact interaction is evaluated by tensor contractions: for each pair $(\alpha,\beta)$, the last two indices of $G$ are contracted with the indices $j_\alpha$ and $j_\beta$ of $T$,
\begin{equation}
    (HT)_{\mathbf j}
    =\Bigl(\sum_{\alpha=1}^{N}j_\alpha+\frac N2\Bigr)T_{\mathbf j}
    +g\sum_{\alpha<\beta}(V_{\alpha\beta}T)_{\mathbf j},
    \qquad
    (V_{12}T)_{j_1j_2j_3\cdots j_N}
    =\sum_{k,l=0}^{n_{\max}}G_{j_1j_2kl}\,T_{kl\,j_3\cdots j_N},
    \label{eq:ED-HT}
\end{equation}
and analogously for every other pair, with all remaining indices as spectators.

The ground state is the lowest eigenpair of the real symmetric matrix $H_B$ of $H$ in the bosonic basis~\eqref{eq:ED-bosonic-basis},
\begin{equation}
    H_B\,\mathbf c=E\,\mathbf c,\qquad
    (H_B)_{\mathbf i\mathbf i'}={}_B\langle\mathbf i|H|\mathbf i'\rangle_B
    =\bigl(U^{\mathsf T}H\,U\bigr)_{\mathbf i\mathbf i'},
    \label{eq:ED-HB}
\end{equation}
where the last equality holds because the columns of $U$ are the basis states~\eqref{eq:ED-bosonic-basis} written in the product basis.
The $\dim\pazocal H_B\times\dim\pazocal H_B$ matrix $H_B$ is never formed; instead, each product $\mathbf c\mapsto H_B\mathbf c=U^{\mathsf T}H\,U\mathbf c$ is evaluated in four steps:
\begin{enumerate}[label=(\roman*),itemsep=0pt,topsep=3pt]
    \item form the symmetric tensor $T=U\mathbf c$;
    \item multiply each entry $T_{\mathbf j}$ by its one-body energy $\sum_\alpha j_\alpha+N/2$;
    \item add the contact term, obtained from the $N(N-1)/2$ tensor contractions $V_{\alpha\beta}T$ of Eq.~\eqref{eq:ED-HT} and multiplied by $g$;
    \item map the result back with $U^{\mathsf T}$.
\end{enumerate}

The lowest eigenpair is computed with an implicitly restarted Lanczos method targeting the smallest algebraic eigenvalue, and the eigenvector is normalized after convergence.
In scans over $g$ at fixed basis, the converged ground state at the previous coupling is used as the starting vector, which reduces the number of iterations without changing the eigenproblem.
The solver tolerance and, in convergence checks, the residual norm $\|H_B\mathbf c-E\mathbf c\|$ serve as numerical diagnostics.

\subsubsection{One-body density matrix}
\label{app:ED-obs}

With $T=U\mathbf c$ for the normalized ground state, the one-body reduced density matrix in the orbital basis is
\begin{equation}
    \rho_{ij}=N\sum_{k_2,\dots,k_N=0}^{n_{\max}}
    T^{*}_{i\,k_2\cdots k_N}\,T_{j\,k_2\cdots k_N},
    \qquad \operatorname{Tr}\rho=N.
    \label{eq:ED-rdm}
\end{equation}
Its eigenvalues are the occupations $\lambda_\ell$, from which the correlation measures defined in the main text are computed.

\subsubsection{Two-body problem in relative coordinates}
\label{app:ED-N2}

For $N=2$ the Hamiltonian separates exactly into center-of-mass and relative motion.
Exploiting this separation makes the two-body calculation an independent test of the HO representation of the contact interaction and of the extrapolation procedure, rather than a smaller instance of the $N\ge3$ calculation.
With the orthonormal coordinates $R=(x_1+x_2)/\sqrt2$ and $r=(x_1-x_2)/\sqrt2$, and $\delta(x_1-x_2)=\delta(r)/\sqrt2$,
\begin{equation}
    H=h_R+h_r+g_{\rm rel}\,\delta(r),
    \qquad g_{\rm rel}=\frac{g}{\sqrt2},
    \label{eq:ED-H2}
\end{equation}
where $h_R$ and $h_r$ are HO Hamiltonians in $R$ and $r$.
The center of mass is in its ground state (energy $1/2$), and the relative ground state is even, so only the orbitals $\varphi_{2k}(r)$ with $k=0,\dots,k_{\max}$ are needed; the relative cutoff is $n^{\rm rel}_{\max}=2k_{\max}$.
In this basis,
\begin{equation}
    (H_{\rm rel})_{kk'}=\Bigl(2k+\frac12\Bigr)\delta_{kk'}
    +g_{\rm rel}\,\varphi_{2k}(0)\,\varphi_{2k'}(0),
    \qquad
    \varphi_{2k}(0)^2=\frac{(2k)!}{\sqrt{\pi}\,4^k(k!)^2},
    \label{eq:ED-Hrel}
\end{equation}
so the contact term is a rank-one matrix.
The total energy is $E=\tfrac12+E_{\rm rel}$, with $E_{\rm rel}$ the lowest eigenvalue of $H_{\rm rel}$.

On the repulsive ground-state branch, the exact relative energy is the solution of
\begin{equation}
    g_{\rm rel}=-2\,
    \frac{\Gamma\bigl(\tfrac34-\tfrac12E_{\rm rel}\bigr)}
    {\Gamma\bigl(\tfrac14-\tfrac12E_{\rm rel}\bigr)},
    \qquad \tfrac12<E_{\rm rel}<\tfrac32,
    \label{eq:ED-exact2}
\end{equation}
which provides a stringent independent check of the basis calculation.

To evaluate the same one-body observables as for $N\ge3$, the relative ground state $\sum_k c_k|2k\rangle_r$ is transformed to the product basis with the one-dimensional Moshinsky transformation~\cite{Moshinsky1959},
\begin{equation}
    |2k\rangle_r|0\rangle_R
    =2^{-k}\sum_{i=0}^{2k}(-1)^i\binom{2k}{i}^{1/2}|i,2k-i\rangle,
    \label{eq:ED-moshinsky}
\end{equation}
which follows from the binomial expansion of $(a_r^\dagger)^{2k}$, with the relative raising operator expressed through the single-particle ones as $a_r^\dagger=(a_1^\dagger-a_2^\dagger)/\sqrt2$, and the resulting symmetric coefficient matrix $T_{ij}$ is used in Eq.~\eqref{eq:ED-rdm}.

The relative cutoff has a different meaning from the one-particle cutoff $n_{\max}$, but the same asymptotic convergence.
For large $k$, $\varphi_{2k}(0)^2\simeq1/(\pi\sqrt k)$, while the energy denominators of the omitted states in the relative resolvent grow as $2k$, so the truncation error scales as $\sum_{k>k_{\max}}k^{-3/2}\propto k_{\max}^{-1/2}$.
The same cutoff dependence is expected for $N\ge3$ (Sec.~\ref{app:ED-pade}).

\subsubsection{Extrapolation to the infinite-basis limit}
\label{app:ED-pade}

The quantities $A\in\{E,1-p_0,S_1,K\}$ are computed for a sequence of cutoffs and extrapolated in the variable
\begin{equation}
    \xi=(n_{\rm cut}+1)^{-1/2},\qquad
    n_{\rm cut}=\begin{cases}
        n^{\rm rel}_{\max}, & N=2,\\[2pt]
        n_{\max}, & N\ge3,
    \end{cases}
    \label{eq:ED-xi}
\end{equation}
so that the infinite-basis limit corresponds to $\xi\to0$.
For $N=2$, this choice follows the two-body asymptotics in Sec.~\ref{app:ED-N2}.
For $N\ge3$, the same pairwise contact cusp persists, giving a $k^{-4}$ high-momentum tail.
Since a one-particle HO cutoff $n_{\max}$ corresponds to a characteristic momentum scale $k_{\max}\propto\sqrt{n_{\max}}$, a leading energy-cutoff correction proportional to $n_{\max}^{-1/2}$ is expected also in the many-body problem.
This is consistent with analyses of contact-interaction basis errors~\cite{Ernst2011,Jeszenszki2018} and with the convergence of the energy in a one-particle 1D HO basis, which was derived analytically for two particles and confirmed numerically for larger particle numbers in Ref.~\cite{Grining2015}.
The argument motivates the extrapolation variable in Eq.~\eqref{eq:ED-xi}; it does not by itself establish the same leading cutoff dependence for all one-body observables.

The stability of the extrapolated limit is assessed by changing the fit window and comparing admissible Pad\'e orders.
The data are fitted with a diagonal Pad\'e approximant of order $[d/d]$, i.e., a ratio of two polynomials in $\xi$ that are both of degree $d$,
\begin{equation}
    A^{[d/d]}(\xi)=\frac{a_0+a_1\xi+\dots+a_d\xi^d}{1+b_1\xi+\dots+b_d\xi^d},
    \qquad
    A_\infty=A^{[d/d]}(0)=a_0.
    \label{eq:ED-pade}
\end{equation}
Its $2d+1$ coefficients are obtained by solving the linearized equations
\begin{equation}
    A(\xi)\bigl(1+b_1\xi+\dots+b_d\xi^d\bigr)\simeq a_0+a_1\xi+\dots+a_d\xi^d
    \label{eq:ED-pade-lin}
\end{equation}
in the least-squares sense over the cutoffs of the fit window; at least $2d+1$ cutoffs are therefore required.
We use $d=3$, i.e., a $[3/3]$ approximant with seven coefficients, for $N=2$, and $d=2$, i.e., a $[2/2]$ approximant with five coefficients, for $N\ge3$.

We verified the choice of $\xi$ directly by solving the $N=2$ problem in the same one-particle product basis, and over the same cutoff windows, as used for $N\ge3$: for $g\in\{5,20,50,100\}$, the error of the energy extrapolated with the $[2/2]$ approximant in the variable of Eq.~\eqref{eq:ED-xi} is at most $4\times10^{-5}$ ($n_{\max}=110$--$140$) and $3\times10^{-4}$ ($n_{\max}=24$--$40$) of the error of the raw energy at the largest cutoff, whereas with the same approximant in $(n_{\max}+1)^{-1}$ it is about $20\%$ of that error.
For $N=3$ and $g=33$, a fit to $n_{\max}=110$--$140$ predicts independently computed energies at $n_{\max}=150$ and $160$ to within $10^{-10}$.
In a few cases the $[2/2]$ fit develops a spurious pole, or a nearly cancelled pole--zero pair, close to the fitted interval; there we use the $[3/3]$ approximant on the same cutoffs, which is pole-free in all such cases.
This affects only one-body measures ($1-p_0$ for $N=3$, $g\ge50$, and $1-p_0$, $S_1$ and $K$ for $N=5$, $g=10$), and the two approximants differ by less than $0.2\%$ except for $N=5$, $g=10$, where they differ by up to $4\%$.
For $N=4$, the few $[2/2]$ fits with a pole contain an almost exactly cancelled pole--zero pair, which does not affect the extrapolated limit; the $[1/1]$, $[2/2]$ and $[3/3]$ limits then agree to within $0.05\%$, and we keep the $[2/2]$ values.
The independent quantities $E$, $1-p_0$, $S_1$ and $K$ are extrapolated separately at each coupling; we then set $S_2=\ln K$ and $p_0=1-(1-p_0)$.
The $N=3$, $4$ and $5$ coupling scans use, respectively, the cutoff windows $n_{\max}=110,115,\ldots,140$; $35,40,50,55,60,65,67,69,70$; and $24,26,\ldots,40$.
For the $N=4$, $g=25$ convergence series, the fit uses $n_{\max}=50,55,60,65,67,69,70$.

\subsubsection{Numerical parameters}
\label{app:ED-params}

Table~\ref{tab:ED-params} summarizes the cutoff conventions and the largest basis sizes used in the convergence studies.
The couplings satisfy $Ng\simeq100$, so that comparably strong correlation regimes are probed for each $N$.
Cutoff values for different $N$ are not directly comparable: for $N=2$, $n^{\rm rel}_{\max}$ is the largest (even) relative quantum number, so $n^{\rm rel}_{\max}=600$ corresponds to only $k_{\max}+1=301$ relative basis states, whereas for $N\ge3$, $n_{\max}$ is the largest one-particle orbital and the basis size is the dimension of the bosonic Hilbert space, Eq.~\eqref{eq:ED-dim}.

\begin{table}[ht]
    \centering
    \caption{Cutoff conventions and largest basis sizes used in the convergence calculations.}
    \label{tab:ED-params}
    \begin{tabular}{clllrl}
        \toprule
        $N$ & $g$ & cutoff definition & largest cutoff & basis size & fit variable \\
        \midrule
        2 & 50 & relative, even HO & $n^{\rm rel}_{\max}=600$ & $301$         & $(n^{\rm rel}_{\max}+1)^{-1/2}$ \\
        3 & 33 & one-particle HO   & $n_{\max}=140$          & $477\,191$     & $(n_{\max}+1)^{-1/2}$ \\
        4 & 25 & one-particle HO   & $n_{\max}=70$           & $1\,150\,626$  & $(n_{\max}+1)^{-1/2}$ \\
        5 & 20 & one-particle HO   & $n_{\max}=40$           & $1\,221\,759$  & $(n_{\max}+1)^{-1/2}$ \\
        \bottomrule
    \end{tabular}
\end{table}

\subsection{Few-body benchmarks}
\label{supp: few-body}

\begin{figure}[!ht]
    \centering
    \includegraphics[width=\textwidth]{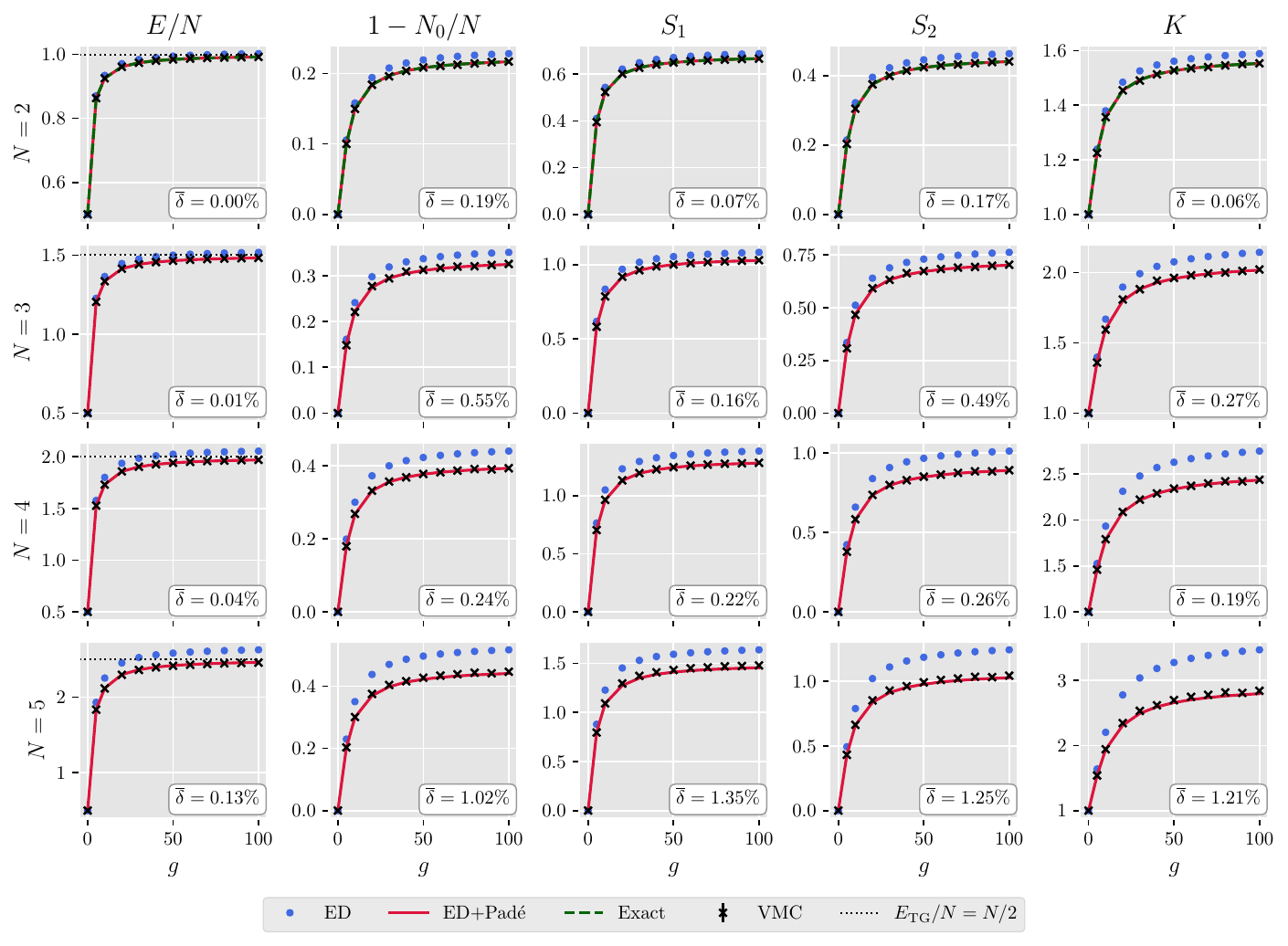}
    \caption{Ground-state properties of the 1D gas with $N=2$--$5$ atoms (rows) as functions of the coupling strength $g$: energy per atom $E/N$, condensate depletion $1-N_0/N=1-p_0$, von Neumann and R\'enyi-2 single-particle entanglement entropies $S_1$ and $S_2$, and effective Schmidt rank $K$ (columns).
    Blue dots: ED at the largest cutoff of Table~\ref{tab:ED-params} ($n^{\rm rel}_{\max}=600$ for $N=2$, and $n_{\max}=140$, $70$ and $40$ for $N=3$, $4$ and $5$); red lines: Pad\'e-extrapolated ED (Sec.~\ref{app:ED-pade}); green dashed lines: exact two-body results; dotted lines: TG energy $E_{\rm TG}/N=N/2$.
    Black crosses: VMC; error bars, from jackknife resampling over 8 Markov chains, lie within the markers.
    The boxes give the relative deviation of VMC from Pad\'e-extrapolated ED, $|A_{\rm VMC}-A_{\rm Pad\acute{e}}|/|A_{\rm Pad\acute{e}}|$ for each quantity $A$, averaged over the couplings $g>0$ and denoted $\overline{\delta}$.}
    \label{fig: small N g series}
\end{figure}

\begin{figure}[!ht]
    \centering
    \includegraphics[width=\textwidth]{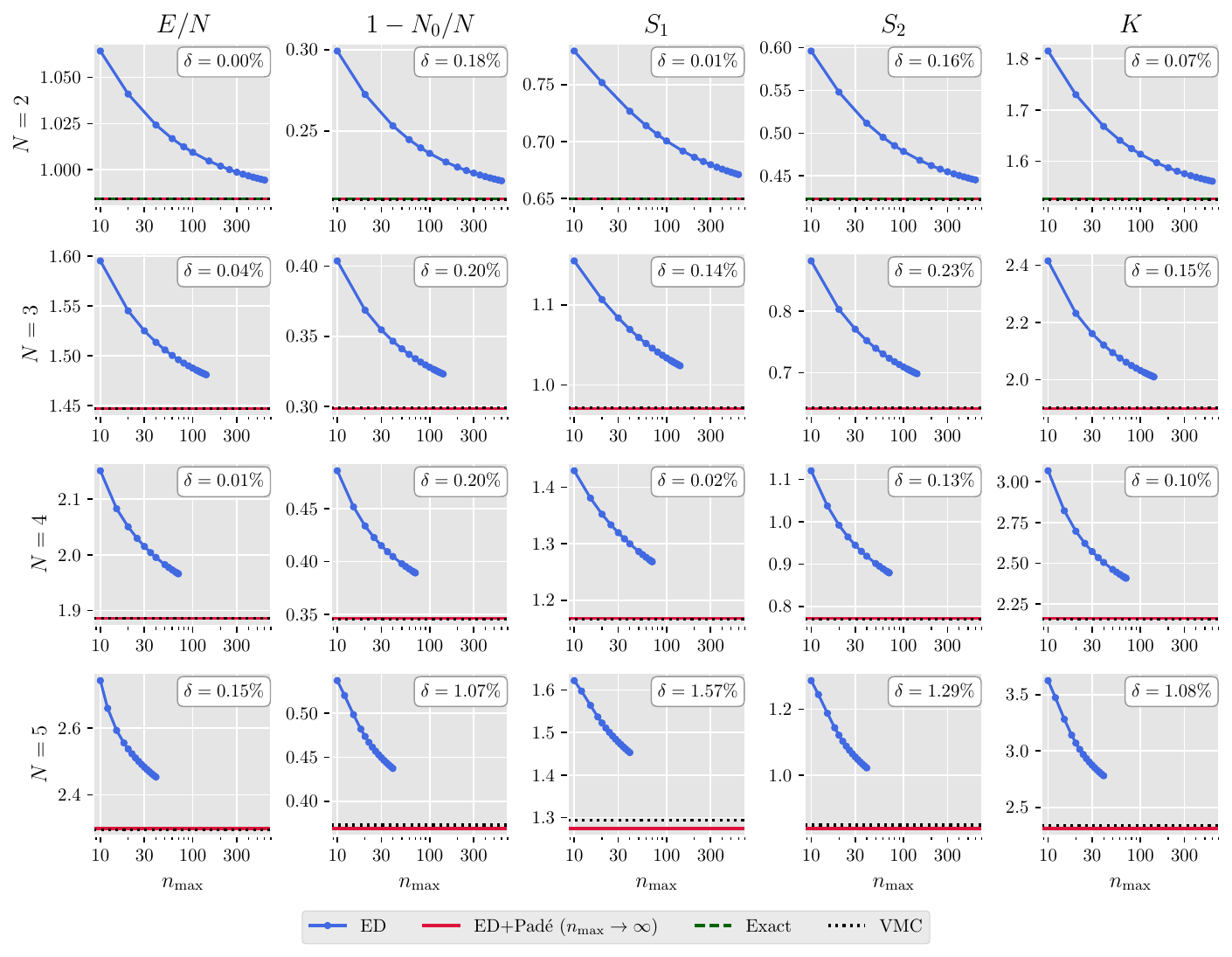}
    \caption{Convergence of ED with the basis cutoff for the 1D gas with $N=2$--$5$ atoms (rows) at $Ng\simeq100$ ($g=50$, $33$, $25$ and $20$; Table~\ref{tab:ED-params}), for the quantities of Fig.~\ref{fig: small N g series} (columns).
    Blue: ED as a function of the relative cutoff $n^{\rm rel}_{\max}$ for $N=2$ and of the one-particle cutoff $n_{\max}$ for $N\ge3$; red lines: Pad\'e extrapolation to $n_{\max}\to\infty$ over the fit windows of Sec.~\ref{app:ED-params}; green dashed lines: exact two-body results; black dotted lines: VMC, taken from Fig.~\ref{M-fig: large N limit} of the main text.
    The boxes give the relative deviation of VMC from Pad\'e-extrapolated ED, $\delta=|A_{\rm VMC}-A_{\rm Pad\acute{e}}|/|A_{\rm Pad\acute{e}}|$.}
    \label{fig: small N nmax}
\end{figure}

In Figs.~\ref{fig: small N g series} and~\ref{fig: small N nmax}, we compare the VMC results for the 1D gas with $N=2$--$5$ atoms to the ED calculations of Sec.~\ref{supp: ED} and, for $N=2$, to the exact two-body solution~\cite{Busch1998}.
Fig.~\ref{fig: small N g series} shows the energy per atom and the correlation measures of the main text as functions of the coupling strength for $g\in\{0,5,10,20,30,\dots,100\}$, while Fig.~\ref{fig: small N nmax} shows the convergence of ED with the basis cutoff at $Ng\simeq100$, i.e., along the scaling trajectory of Fig.~\ref{M-fig: large N limit} in the main text.

For $N=2$, the pair-product ansatz contains the exact two-body ground state, and the VMC energy coincides with the exact result at all couplings.
ED, by contrast, converges slowly with the basis cutoff because of the cusp of the wavefunction at the contact~\cite{Ernst2011,Jeszenszki2018}: at $n^{\rm rel}_{\max}=600$, the energy and correlation measures still deviate from the exact values by up to $5.5\%$, and only the Pad\'e extrapolation, which uses the exact asymptotics in the relative cutoff derived in Sec.~\ref{app:ED-N2}, reproduces them to within $0.04\%$ (Fig.~\ref{fig: small N nmax}, top row).
The VMC correlation measures agree with the exact ones to within $\overline{\delta}\le0.2\%$ across the coupling grid, and to within $0.6\%$ at every coupling.
Since the trial state is exact for $N=2$, these small residual deviations are due solely to the Monte Carlo estimate of the OBDM.

For $3\le N\le5$, VMC and ED are in excellent quantitative agreement over the entire range of couplings.
The VMC energies lie $1.8$--$6.5\%$ below the ED energies at all accessible cutoffs, and agree with the extrapolated ED energies to within $0.07\%$ for $N=3$ and $4$, comparable to the statistical error.
Since the VMC energy is a rigorous variational upper bound to the exact ground-state energy, VMC, with a trial state of only two variational parameters, thus provides a tighter bound on the ground-state energy than ED in bases of up to $1.2\times10^6$ states (Table~\ref{tab:ED-params}).
For $N=5$, the VMC energies lie $0.13$--$0.15\%$ below the extrapolated ED energies for $g\ge10$, well outside the statistical error; as VMC is an upper bound, this indicates a residual extrapolation error of at least this size at the smallest attainable cutoffs.
The correlation measures agree to within $\overline{\delta}\le0.6\%$ for $N\le4$ and $\overline{\delta}\le1.4\%$ for $N=5$.

Taken together, these benchmarks establish our VMC method as highly accurate in the few-body regime, which reproduces the exact two-body solution and yields lower variational energies than finite-basis ED for $N\ge3$, in agreement with the extrapolated ED to within $0.15\%$, together with closely matching correlation measures.
At the same time, the cost of evaluating the local energy grows only as $O(N^2)$ per sample (Sec.~\ref{supp: kinetic}), compared with the combinatorial growth of the ED basis, Eq.~\eqref{eq:ED-dim}.

\bibliography{references.bib}